\documentclass[11pt,draftcls,onecolumn]{IEEEtran}
\usepackage{graphicx}
\usepackage{amsmath}
\usepackage{amsfonts}
\usepackage{amsthm}
\usepackage{cite}
\usepackage{subfigure}

\newcommand{\bl}{\left(}
\newcommand{\br}{\right)}
\newcommand{\blc}{\left\{}
\newcommand{\brc}{\right\}}

\newcommand{\ba}{{\bf a}}

\newcommand{\bn}{{\bf n}}

\newcommand{\hbbF}{{\widehat{\bf F}}}

\newcommand{\bx}{{\bf x}}

\newcommand{\bbsg}{{\bf \Sigma}}
\newcommand{\bbla}{\mathbf \Lambda}
\newcommand{\bbi}{{\bf I}}

\newcommand{\bbd}{{{\bf D}}}

\newcommand{\bbs}{{\bf S}}

\newcommand{\bbh}{{\bf H}}

\newcommand{\bbx}{{\bf X}}

\newcommand{\hs}{\widehat{\bbs}}

\newcommand{\hsg}{\widehat{\bbsg}}

\newcommand{\bbq}{{\bf Q}}
\newcommand{\bbm}{{\bf M}}

\newcommand{\bba}{{\bf A}}

\newcommand{\bbb}{{\bf B}}

\newcommand{\tr}{\mathrm{Tr}}

\newtheorem{lem}{Lemma}
\newtheorem{thm}{Theorem}
\newtheorem{definition}{Definition}

\def\({\left (}
\def\){\right)}
\def\l{\left}
\def\r{\right}
\def\det{{\mathrm{det}}}
\renewcommand{\vec}[1]{\mathbf{#1}}

\title{Shrinkage Algorithms for MMSE Covariance Estimation}
\author{Yilun~Chen,
        Ami~Wiesel,
        Yonina~C.~Eldar
        and Alfred~O.~Hero~III
\thanks{Y. Chen, A. Wiesel and A. O. Hero are with the Department
of Electrical Engineering and Computer Science, University of
Michigan, Ann Arbor, MI 48109, USA. Tel: (734) 763-0564, Fax:
(734) 763-8041.
Emails: \{yilun,amiw,hero\}@umich.edu.}
\thanks{Y. C. Eldar is with the Technion - Israel Institute of Technology,
Haifa, Israel 32000. Email: yonina@ee.technion.ac.il.}
\thanks{This research was partially supported by the AFOSR
grant FA9550-06-1-0324 and NSF grant CCF 0830490. The work of A.
Wiesel was supported by a Marie Curie Outgoing International
Fellowship within the 7th European Community Framework
Programme.}}
\begin{document}
\maketitle

\begin{abstract}
We address covariance estimation in the sense of minimum
mean-squared error (MMSE) for Gaussian samples. Specifically, we
consider shrinkage methods which are suitable for high dimensional
problems with a small number of samples (large $p$ small $n$).
First, we improve on the Ledoit-Wolf (LW) method by conditioning
on a sufficient statistic. By the Rao-Blackwell theorem, this
yields a new estimator called RBLW, whose mean-squared error
dominates that of LW for Gaussian variables. Second, to further
reduce the estimation error, we propose an iterative approach
which approximates the clairvoyant shrinkage estimator.
Convergence of this iterative method is established and a closed
form expression for the limit is determined, which is referred to
as the oracle approximating shrinkage (OAS) estimator. Both RBLW
and OAS estimators have simple expressions and are easily
implemented. Although the two methods are developed from different
persepctives, their structure is identical up to specified
constants. The RBLW estimator provably dominates the LW method.
Numerical simulations demonstrate that the OAS approach can
perform even better than RBLW, especially when $n$ is much less
than $p$. We also demonstrate the performance of these techniques
in the context of adaptive beamforming.
\end{abstract}


\section{Introduction}
Covariance matrix estimation is a fundamental problem in signal
processing and related fields. Many applications varying from
array processing \cite{Abrahamsson2007} to functional genomics
\cite{Schafer2005} rely on accurately estimated covariance
matrices. In recent years, estimation of high dimensional $p
\times p$ covariance matrices under small sample size $n$ has
attracted considerable interest. Examples include classification
on gene expression from microarray data, financial forecasting,
spectroscopic imaging, brain activation mapping from fMRI and many
others. Standard estimation methods perform poorly in these large
$p$ small $n$ settings. This is the main motivation for this work.

The sample covariance is a common estimate for the unknown
covariance matrix. When it is invertible, the sample covariance
coincides with the classical maximum likelihood estimate. However,
while it is an unbiased estimator, it does not minimize the
mean-squared error (MSE). Indeed, Stein demonstrated that superior
performance may be obtained by shrinking the sample covariance
\cite{Stein1961,Stein1975}. Since then, many shrinkage estimators
have been proposed under different performance measures. For
example, Haff \cite{Haff} introduced an estimator inspired by the
empirical Bayes approach. Dey and Srinivasan \cite{Dey} derived a
minimax estimator under Stein's entropy loss function. Yang and
Berger \cite{Yang} obtained expressions for Bayesian estimators
under a class of priors for the covariance. These works addressed
the case of invertible sample covariance when $n \ge p$. Recently,
Ledoit and Wolf (LW) proposed a shrinkage estimator for the case
$n < p$ which asymptotically minimizes the MSE\cite{Ledoit2004}.
The LW estimator is well conditioned for small sample sizes and
can thus be applied to high dimensional problems. In contrast to
previous approaches, they show that performance advantages are
distribution-free and not restricted to Gaussian assumptions.

In this paper, we show that the LW estimator can be significantly
improved when the samples are in fact Gaussian. Specifically, we
develop two new estimation techniques that result from different
considerations. The first follows from the Rao-Blackwell theorem,
while the second is an application of the ideas of
\cite{Eldar2008} to covariance estimation.

We begin by providing a closed form expression for the optimal
clairvoyant shrinkage estimator under an MSE loss criteria. This
estimator is an explicit function of the unknown covariance matrix
that can be used as an oracle performance bound. Our first
estimator is obtained by applying the well-known Rao-Blackwell
theorem \cite{Trees} to the LW method, and is therefore denoted by
RBLW. Using several nontrivial Haar integral computations, we
obtain a simple closed form solution which provably dominates the
LW method in terms of MSE. We then introduce an iterative
shrinkage estimator which tries to approximate the oracle. This
approach follows the methodology developed in \cite{Eldar2008} for
the case of linear regression. Beginning with an initial naive
choice, each iteration is defined as the oracle solution when the
unknown covariance is replaced by its estimate obtained in the
previous iteration. Remarkably, a closed form expression can be
determined for the limit of these iterations. We refer to the
limit as the oracle approximating shrinkage (OAS) estimator.

The OAS and RBLW solutions have similar structure that is related
to a sphericity test as discussed in
\cite{John,Srivastava,Ledoit3}. Both OAS and RBLW estimators are
intuitive, easy to compute and perform well with finite sample
size. The RBLW technique provably dominates LW. Numerical results
demonstrate that for small sample sizes, the OAS estimator is
superior to both the RBLW and the LW methods.

To illustrate the proposed covariance estimators we apply them to
problems of time series analysis and array signal processing.
Specifically, in the context of time series analysis we establish
performance advantages of OAS and RBLW to LW for covariance
estimation in autoregressive models and in fractional Brownian
motion models, respectively. In the context of beamforming, we
show that RBLW and OAS can be used to significantly improve the
Capon beamformer. In \cite{Abrahamsson2007} a multitude of
covariance matrix estimators were implemented in Capon
beamformers, and the authors reported that the LW approach
substantially improves performance as compared to other methods.
We show here that even better performance can be achieved by using
the techniques introduced in this paper.

The paper is organized as follows. Section 2 formulates the
problem. Section 3 introduces the oracle estimator together with
the RBLW and OAS methods. Section 4 represents numerical
simulation results and applications in adaptive beamforming.
Section 5 summarizes our principal conclusions. The proofs of
theorems and lemmas are provided in the Appendix.

\emph{Notation}: In the following, we depict vectors in lowercase
boldface letters and matrices in uppercase boldface letters.
$\(\cdot\)^T$ and $\(\cdot\)^H$ denote the transpose and the
conjugate transpose, respectively. ${\tr}\(\cdot\)$,
$\l\|\cdot\r\|_F$, and $\text{det}\(\cdot\)$ are the trace, the
Frobenius norm, and the determinant of a matrix, respectively.
Finally, $\bba \prec \bbb$ means that the matrix $\bbb-\bba$ is
positive definite, and $\bba \succ \bbb$ means that the matrix
$\bba-\bbb$ is positive definite.

\section{Problem formulation}

Let $\l\{\vec x_i\r\}_{i=1}^n$ be a sample of independent
identical distributed (i.i.d.) $p$-dimensional Gaussian vectors
with zero mean and covariance $\bbsg$. We do not assume $n \ge p$.
Our goal is to find an estimator $\hsg\(\{\vec x_i\}_{i=1}^n\)$
which minimizes the MSE:
\begin{equation}\label{mse}
E\blc\l\|\hsg\(\{\vec x_i\}_{i=1}^n\)-\bbsg\r\|_F^2\brc.
\end{equation}

It is difficult to compute the MSE of $\hsg\(\{\vec
x_i\}_{i=1}^n\)$ without additional constraints and therefore we
restrict ourselves to a specific class of estimators that employ
shrinkage \cite{Stein1956,Ledoit2003}. The unstructured classical
estimator of $\bbsg$ is the sample covariance $\hs$ defined as
\begin{equation}
    \label{eq:hat_S}
    \hs = \frac{1}{n} \sum_{i=1}^n \vec x_i \vec x_i^T.
\end{equation}
This estimator is unbiased $E\{\hs\} = \bbsg$, and is also the
maximum likelihood solution if $n \ge p$. However, it does not
necessarily achieve low MSE due to its high variance and is
usually ill-posed for large $p$ problems. On the other hand, we
may consider a naive but most well-conditioned estimate for
$\bbsg$:
\begin{equation}
  \label{eq:hat_F}
  \hbbF = \frac{\tr\(\hs\)}{p} \bbi.
\end{equation}
This ``structured" estimate will result in reduced variance with
the expense of increasing the bias. A reasonable tradeoff between
low bias and low variance is achieved by shrinkage of $\hs$
towards $\hbbF$, resulting in the following class of estimators:
\begin{equation}
 \label{eq:shrinkage_def}
\hsg = (1-\hat\rho)\hs + \hat\rho\hbbF.
\end{equation}
The estimator $\hsg$ is characterized by the shrinkage coefficient
$\hat \rho$, which is a parameter between 0 and 1 and can be a
function of the observations $\l\{\vec x_i\r\}_{i=1}^n$. The
matrix $\hbbF$ is referred to as the shrinkage
target.\footnote{The convex combination
 in (\ref{eq:shrinkage_def}) can be generalized to the linear combination of
$\hs$ and $\hbbF$. The reader is referred to \cite{Stoica2008} for
further discussion.}

Throughout the paper, we restrict our attention to shrinkage
estimates of the form (\ref{eq:shrinkage_def}). Our goal is to
find a shrinkage coefficient $\hat \rho$ that minimizes the MSE
(\ref{mse}). As we show in the next section, the optimal
$\hat\rho$ minimizing the MSE depends in general on the unknown
$\bbsg$ and therefore in general cannot be implemented. Instead,
we propose two different approaches to approximate the optimal
shrinkage coefficient.

\section{Shrinkage algorithms}
\subsection{The Oracle estimator}
We begin by deriving a clairvoyant oracle estimator that uses an
optimal nonrandom coefficient to minimize the mean-squared error.
In the following subsections we will show how to approximate the
oracle using implementable data-driven methods.

The oracle estimate $\hsg_O$ is the solution to
\begin{equation}
    \label{eq:oracle_def}
\begin{array}{*{20}c}
   {\mathop {\min }\limits_{\rho} } & {E\blc {\l\|\hsg_O  - \bbsg \r\|_F^2 } \brc} \\
   \text{s.t.} & {\hsg_O  = \( {1 - \rho } \)\hs} + \rho  \hbbF
\end{array}.
\end{equation}
The optimal parameter $\rho_O$ is provided in the following
theorem.

\begin{thm}
  \label{thm:oracle}
Let $\hs$ be the sample covariance of a set of $p$-dimensional
vectors  $\l\{\vec x_i\r\}_{i=1}^n$. If  $\l\{\vec
x_i\r\}_{i=1}^n$ are i.i.d. Gaussian vectors with covariance
$\bbsg$, then the solution to (\ref{eq:oracle_def}) is
\begin{align}
    \label{eq:oracle_rho1}
    \rho_O & = \frac{E\blc\tr \(\(\bbsg-\hs\)\(\hbbF - \hs\)\)\brc}
    {E\blc\l\|\hs-\hbbF\r\|^2_F\brc} \\
    \label{eq:oracle_rho2}
    & = \frac{\(1-2/p\){\mathrm{Tr}\left( {\bbsg ^2 }
    \right) + \mathrm{Tr}^2 \left( \bbsg  \right)}}{(n+1-2/p)\mathrm{Tr}(\bbsg^2) +
    (1-n/p)~
    \mathrm{Tr}^2(\bbsg)}.
\end{align}
\end{thm}
\begin{proof}
Equation (\ref{eq:oracle_rho1}) was established in
\cite{Ledoit2003} for any distribution of $\blc\vec
x_i\brc_{i=1}^n$. Under the additional Gaussian assumption,
(\ref{eq:oracle_rho2}) can be obtained from straightforward
evaluation of the expectations:
\begin{equation}
\begin{aligned}
    \label{eq:thm_oracle_1}
    E & \blc\tr\(\(\bbsg-\hs\)\(\hbbF - \hs\)\)\brc
     = \frac{\tr\(\bbsg\)}{p}E\blc\tr\(\hs\)\brc \\
     & \quad \quad - \frac{E\blc\tr^2\(\hs\)\brc}{p} - E\blc\tr\(\bbsg \hs\)\brc +
     E\blc\tr\(\hs^2\)\brc,
\end{aligned}
\end{equation}
and
\begin{equation}
\begin{aligned}
    \label{eq:thm_oracle_2}
    E & \blc\l\|\hs-\hbbF\r\|^2_F\brc \\ & = E\blc\tr\(\hs^2\)\brc -
    2E\blc\tr\(\hs \hbbF\)\brc + E\blc\tr\(\hbbF^2\)\brc \\
    & = E\blc\tr\(\hs^2\)\brc -\frac{E\blc\tr^2\(\hs\)\brc}{p}.
\end{aligned}
\end{equation}
Equation (\ref{eq:oracle_rho2}) is a result of using the following
identities \cite{Letac}:
\begin{equation}
E\blc {\tr\left( \hs \right)} \brc = \tr\left( \bbsg \right),
\end{equation}
\begin{equation}
E\blc {\tr\left( {\hs^2 } \right)} \brc = \frac{n+1}{n}\tr\left(
  {\bbsg ^2 } \right) + \frac{1}{n}\tr^2 \left( \bbsg  \right),
\end{equation}
and
\begin{equation}
E\blc {\tr^2 \left( \hs \right)} \brc = \tr^2 \left( \bbsg \right)
+ \frac{2}{n}\tr\left( {\bbsg ^2 } \right).
\end{equation}
\end{proof}
Note that (\ref{eq:oracle_rho1}) specifies the optimal shrinkage
coefficient for any sample distribution while
$(\ref{eq:oracle_rho2})$ only holds for the Gaussian distribution.

\subsection{The Rao-Blackwell Ledoit-Wolf (RBLW) estimator}
The oracle estimator defined by (\ref{eq:oracle_def}) is optimal
but cannot be implemented, since the solution specified by both
(\ref{eq:oracle_rho1}) and (\ref{eq:oracle_rho2}) depends on the
unknown $\bbsg$. Without any knowledge of the sample distribution,
Ledoit and Wolf \cite{Ledoit2003, Ledoit2004} proposed to
approximate the oracle using the following consistent estimate of
(\ref{eq:oracle_rho1}):
\begin{equation}
    \label{eq:rho_L}
{\hat \rho_{LW} = {\frac{{\sum\limits_{i = 1}^n {\l\|\vec x_i \vec
x_i^T
            - \hs\r\|_F^2 } }}{{n^2 \left[\tr\(\hs^2\) - \tr^2\(\hs\)/p\right]
            }}}
  }.
\end{equation}
They then proved that when both $n,p \rightarrow \infty$ and $p/n
\rightarrow c$, $0 < c < \infty$, (\ref{eq:rho_L}) converges to
(\ref{eq:oracle_rho1}) in probability regardless of the sample
distribution. The LW estimator $\hsg_{LW}$ is then defined by
plugging $\hat \rho_{LW}$ into (\ref{eq:shrinkage_def}). In
\cite{Ledoit2004} Ledoit and Wolf also showed  that the optimal
$\rho_O$ (\ref{eq:oracle_rho1}) is always between 0 and 1. To
further improve the performance,
 they suggested using a modified shrinkage parameter
 \begin{equation}
   \label{eq:rho_LW_}
   \hat\rho_{LW}^* = \min\(\hat \rho_{LW},1\)
 \end{equation}
instead of $\hat \rho_{LW}$.

The Rao-Blackwell LW (RBLW) estimator described below provably
improves on the LW method under the Gaussian model. The motivation
for the RBLW originates from the fact that under the Gaussian
assumption on $\blc\bx_i\brc_{i=1}^n$, a sufficient statistic for
estimating $\bbsg$ is the sample covariance $\hs$. Intuitively,
the LW estimator is a function of not only $\hs$ but other
statistics and therefore, by the Rao-Blackwell theorem, can be
improved. Specifically, the Rao-Blackwell theorem \cite{Trees}
states that if $g(X)$ is an estimator of a parameter $\theta$,
then the conditional expectation of $g(X)$ given $T(X)$, where $T$
is a sufficient statistic, is never worse than the original
estimator $g(X)$ under any convex loss criterion. Applying the
Rao-Blackwell theorem to the LW estimator yields the following
result.

\begin{thm}
    \label{thm:RBLW}
Let $\{\vec x_i\}_{i=1}^n$ be independent $p$-dimensional Gaussian
vectors with covariance $\bbsg$, and let $\hs$ be the sample
covariance of $\{\vec x_i\}_{i=1}^n$. The conditioned expectation
of the LW covariance estimator is
\begin{eqnarray}
    \label{eq:RBLW}
    \hsg_{RBLW} &=& E\l[\hsg_{LW}\l|\hs\r.\r]\\
    &=&(1-\hat\rho_{RBLW})\hs + \hat \rho_{RBLW}
    \hbbF
\end{eqnarray}
where
\begin{equation}
    \label{eq:rhoR_def}
\hat \rho _{RBLW}  = \frac{{(n - 2)/n \cdot \tr\( {\hs^2 } \) +
\tr^2 \(
      {\hs} \)}}{{(n+2)\left[ {\tr\(\hs^2 \) - \tr^2 \(\hs\)/p} \right]}}.
\end{equation}
This estimator satisfies
\begin{equation}
  E\blc {\l\|\hsg_{RBLW}  - \bbsg \r\|_F^2 } \brc \le E\blc
    {\l\|\hsg_{LW}  - \bbsg \r\|_F^2 } \brc,
\end{equation}
for every $\bbsg$.
\end{thm}

The proof of Theorem \ref{thm:RBLW} is given in the Appendix.

Similarly to the LW estimator, we propose the modification
\begin{equation}
  \label{eq:rho_RBLW_}
  \hat \rho_{RBLW}^* = \min\(\hat \rho_{RBLW},1\)
\end{equation}
instead of $\hat \rho_{RBLW}$.

\subsection{The Oracle-Approximating Shrinkage (OAS) estimator}

The basic idea of the LW estimator is to asymptotically
approximate the oracle, which is designed for large sample size.
For a large number of samples the LW asymptotically achieves the
minimum MSE with respect to shrinkage estimators. Clearly, the
RBLW also inherits this property. However, for very small $n$,
which is often the case of interest, there is no guarantee that
such optimality still holds. To illustrate this point, consider
the extreme example when only one sample is available. For $n = 1$
we have both $\hat\rho^*_{LW} = 1$ and $\hat\rho^*_{RBLW} = 1$,
which indicates that $\hsg_{LW} = \hsg_{RBLW} = \hs$. This however
contradicts our expectations since if a single sample is
available, it is more reasonable to expect more confidence to be
put on the more parsimonious $\hbbF$ rather than $\hs$.

In this section, we consider an alternative approach to
approximate the oracle estimator based on \cite{Eldar2008}. In
(\ref{eq:oracle_rho2}), we obtained a closed-form formula of the
oracle estimator under Gaussian assumptions. The idea behind the
OAS is to approximate this oracle via an iterative procedure. We
initialize the iterations with an initial guess of $\bbsg$ and
iteratively refine it. The initial guess $\hsg_0$ might be the
sample covariance, the RBLW estimate or any other symmetric
non-negative definite estimator. We replace $\bbsg$ in the oracle
solution by $\hsg_0$ yielding $\hsg_1$, which in turn generates
$\hsg_2$ through our proposed iteration. The iteration process is
continued until convergence. The limit, denoted as $\hsg_{OAS}$,
is the OAS solution. Specifically, the proposed iteration is,
\begin{align}
    \label{eq:iteration_rho}
    &\hat{\rho}_{j+1}  = \frac{{(1-2/p)\tr\( {\hsg_j\hs} \)
        + \tr^2 \( \hsg_j  \)
        }}{(n+1-2/p)\tr\(\hsg_j \hs\)
        + (1-n/p)~\tr^2\(\hsg_j\)},\\
    \label{eq:iteration_Sigma}
    &\hsg_{j+1} = (1-\hat{\rho}_{j+1})\hs + \hat{\rho}_{j+1} \hbbF.
\end{align}
Comparing with (\ref{eq:oracle_rho2}), notice that in
(\ref{eq:iteration_rho}) $\tr(\bbsg)$ and $\tr(\bbsg^2)$ are
replaced by $\tr(\hsg_j)$ and $\tr(\hsg_j \hs)$, respectively.
Here $\tr(\hsg_j \hs)$ is used instead of $\tr(\hsg_j^2)$ since
the latter would always force $\hat\rho_j$ to converge to 1 while
the former leads to a more meaningful limiting value.
\begin{thm}
  \label{thm:OAS}
For any initial guess $\hat\rho_0$ that is between $0$ and $1$,
the iterations specified by (\ref{eq:iteration_rho}),
(\ref{eq:iteration_Sigma}) converge as $j \rightarrow \infty$ to
the following estimate:
\begin{align}
    \label{eq:iteration_Sigma_converged}
    &\hsg_{OAS} = (1-\hat{\rho}_{OAS}^*)\hs +
    \hat\rho_{OAS}^* \hbbF,
\end{align}
where
\begin{equation}
    \label{eq:iteration_rho_converged}
    \hat{\rho}_{OAS}^*  = \min \(
\frac{{(1-2/p)\tr\( {\hs^2 } \) + \tr^2 \( {\hs}
    \)}}{{(n+1-2/p)\left[ {\tr\(\hs^2 \) - \tr^2 \(\hs\)/p}
    \right]}}, 1 \).
\end{equation}
In addition, $0<\hat\rho_{OAS}^*\le 1$.
\end{thm}
\begin{proof}
Plugging in $\hsg_{j}$ from (\ref{eq:iteration_Sigma}) into
(\ref{eq:iteration_rho}) and simplifying yields
\begin{equation}
    \label{eq:proof2_iteration_rho_j}
    \hat\rho_{j+1} = \frac{1-(1-2/p)\hat\phi\hat\rho_j}
    {1+n\hat\phi-(n+1-2/p)\hat\phi\hat\rho_j},
\end{equation}
where
\begin{equation}
    \label{eq:proof2_phi}
  \hat{\phi} = \frac{\tr\(\hs^2\)-
    \tr^2\(\hs\)/p}{\tr\(\hs^2\)+
    \tr^2\(\hs\)}.
\end{equation}
Since $\tr(\hs^2) \ge \tr^2(\hs)/p$, $0 \le \hat\phi<1$. Using a
simple change of variables
\begin{equation}\label{eq:b_def}
    \hat b_j = \frac{1}{1 - (n+1-2/p)\hat{\phi}\hat
    \rho _j},
\end{equation}
(\ref{eq:proof2_iteration_rho_j}) is equivalent to the following
geometric series
\begin{equation}
\hat b_{j+1} = \frac{{n\hat{\phi} }}{{1 - (1-2/p)\hat{\phi} }}\hat
b_j + \frac{1}{{1 - (1-2/p)\hat{\phi} }}.
\end{equation}
It is easy to see that
\begin{equation}
\label{eq:converged} \mathop {\lim }\limits_{j \to \infty } \hat
b_j =
  \left \{
\begin{aligned}
    & \infty, & \quad  \frac{n\hat \phi}{1-(1-2/p)\hat
    \phi} \ge 1 \\
    & \frac{1}{1-(n+1-2/p)\hat\phi}, & \quad \frac{n\hat \phi}{1-(1-2/p)\hat
    \phi} < 1
\end{aligned}
 \right..
\end{equation}
Therefore $\hat\rho_j$ also converges as $j \rightarrow \infty$
and $\hat\rho_{OAS}^*$ is given by
\begin{equation}
    \label{eq:proof2_rho_I}
\hat \rho_{OAS}^* = \mathop {\lim }\limits_{j \to \infty } \hat
\rho_j = \left\{ {
    \begin{aligned}
    & {\frac{1}{(n+1-2/p)\hat{\phi}}} & {(n+1-2/p)}{\hat{\phi}}  > 1  \\
    & 1 & {(n+1-2/p)}{\hat{\phi}}  \le 1  \\
    \end{aligned}
} \right. .
\end{equation}

We can write (\ref{eq:proof2_rho_I}) equivalently as
\begin{equation}
\hat \rho _{OAS}^*  = \min \left( {\frac{1}{{(n+1-2/p)
        \hat \phi }},1} \right).
\end{equation}
Equation (\ref{eq:iteration_rho_converged}) is obtained by
substituting (\ref{eq:proof2_phi}) into (\ref{eq:proof2_rho_I}).
\end{proof}

Note that (\ref{eq:proof2_rho_I}) $\hat\rho_{OAS}^*$ is naturally
bounded within $[0,1]$. This is different from $\hat\rho_{LW}^*$
and $\hat\rho_{RBLW}^*$, where the $[0,1]$ constraint is imposed
in an ad hoc fashion.

\subsection{Shrinkage and sphericity statistics}
We now turn to theoretical comparisons between RBLW and OAS. The
only difference is in their shrinkage coefficients. Although
derived from distinct approaches, it is easy to see that $\hat
\rho_{OAS}^*$ shares the same structure as $\hat \rho_{RBLW}^*$.
In fact, they can both be expressed as the parameterized function
\begin{equation}
    \label{eq:rho_E}
    \hat\rho_{E}^* = \min\(\alpha + \frac{\beta}{\hat
    U},1\)
\end{equation}
with $\hat U$ defined as
\begin{equation}
\hat U = \frac{1}{p-1}
  \(\frac{p\cdot \tr\(\hs^2\)}{\tr^2\(\hs\)} -1 \).
\end{equation}
For $\hat\rho^*_E=\hat\rho^*_{OAS}$, $\alpha$ and $\beta$ of
(\ref{eq:rho_E}) are given by
\begin{equation}
\begin{aligned}
  \alpha &= \alpha_{OAS} = \frac{1}{n+1-2/p}\\
  \beta &= \beta_{OAS} =
  \frac{p+1}{(n+1-2/p)(p-1)}
\end{aligned},
\end{equation}
while for $\hat\rho_E^* = \hat\rho_{RBLW}^*$:
\begin{equation}
\begin{aligned}
  \alpha &= \alpha_{RBLW} = \frac{n-2}{n(n+2)}\\
  \beta &= \beta_{RBLW} =
  \frac{(p+1)n-2}{n(n+2)(p-1)}
  \end{aligned}.
\end{equation}

Thus the only difference between $\hat \rho_{OAS}^*$ and $\hat
\rho_{RBLW}^*$ is the choice of $\alpha$ and $\beta$. The
statistic $\hat U$ arises in tests of sphericity of $\bbsg$
\cite{Srivastava, Ledoit3}, \emph{i.e.}, testing whether or not
$\bbsg$ is a scaled identity matrix. In particular, $\hat U$ is
the locally most powerful invariant test statistic for sphericity
under orthogonal transformations \cite{John}. The smaller the
value of $\hat U$, the more likely that $\bbsg$ is proportional to
an identity matrix $\bbi$. Similarly, in our shrinkage algorithms,
the smaller the value of $\hat U$, the more shrinkage occurs in
$\hsg _{OAS}$ and $\hsg_{RBLW}$.

\section{Numerical Simulations}
In this section we implement and test the proposed covariance
estimators. We first compare the estimated MSE of the RBLW and OAS
techniques with the LW method. We then consider their application
to the problem of adaptive beamforming, and show that they lead to
improved performance of Capon beamformers.

\subsection{MSE Comparison}
To test the MSE of the covariance estimators we designed two sets
of experiments with different shapes of $\bbsg$. Such covariance
matrices have been used to study covariance estimators in
\cite{Bickel2008}. We use (\ref{eq:rho_LW_}), (\ref{eq:rho_RBLW_})
and (\ref{eq:iteration_rho_converged}) to calculate the shrinkage
coefficients for the LW, the RBLW and the OAS estimators. For
comparison, the oracle estimator (\ref{eq:oracle_def}) uses the
true $\bbsg$ and is included as a benchmark lower bound on MSE for
comparison. For all simulations, we set $p=100$ and let $n$ range
from $6$ to $30$. Each simulation is repeated 5000 times and the
MSE and shrinkage coefficients are plotted as a function of $n$.
The 95\% confidence intervals of the MSE and shrinkage
coefficients were found to be smaller than the marker size and are
omitted in the figures.

\begin{figure}[htbp]
  \centering
 \subfigure[MSE]{
    \label{fig:1:a} 
    \includegraphics[width=.5\textwidth]{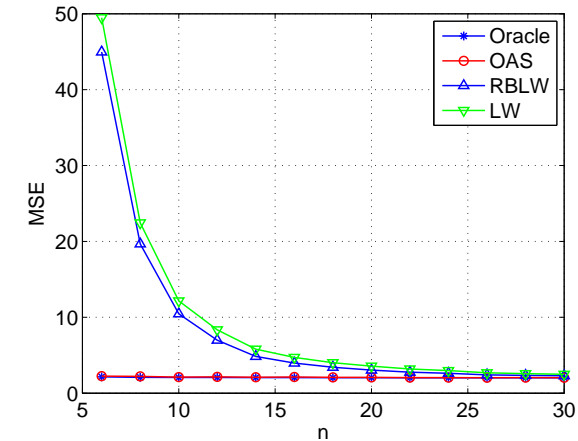}}
   \subfigure[Shrinkage coefficients]{
    \label{fig:1:b} 
    \includegraphics[width=.5\textwidth]{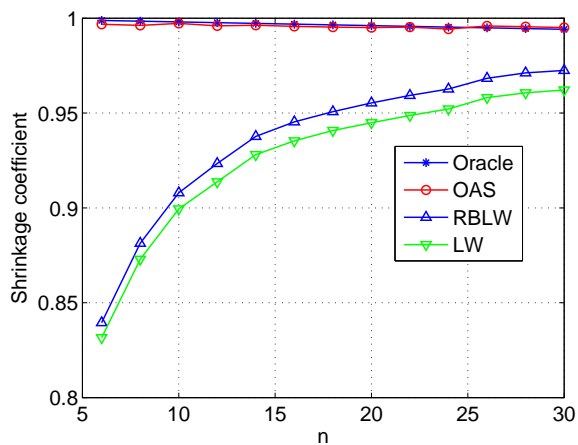}}
  \caption{AR(1) process: Comparison of covariance estimators when $p
    = 100$, $r=0.1$.}
 \label{fig:1}
\end{figure}

\begin{figure}[htbp]
  \centering
 \subfigure[MSE]{
    \label{fig:2:a} 
    \includegraphics[width=.5\textwidth]{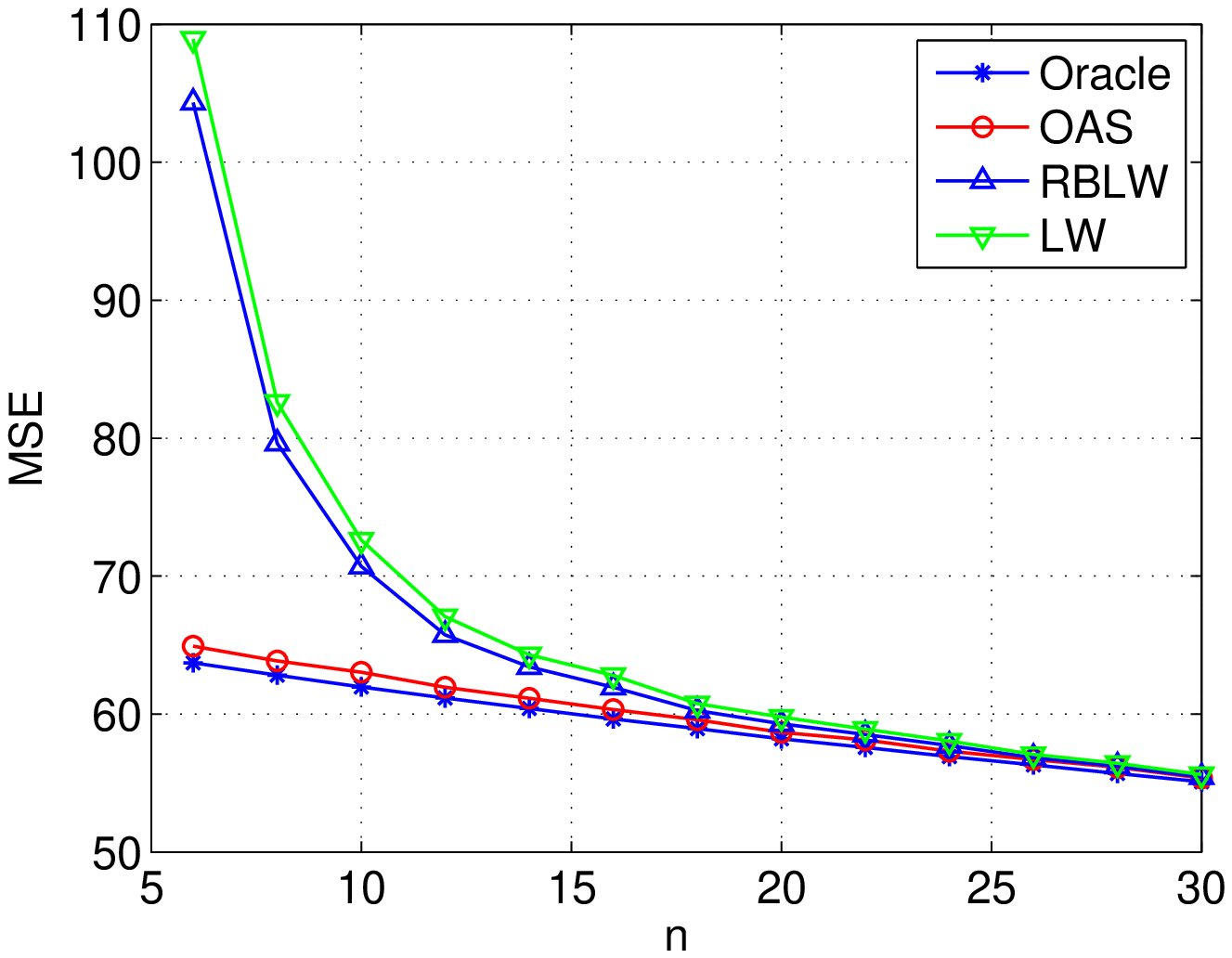}}
   \subfigure[Shrinkage coefficients]{
    \label{fig:2:b} 
    \includegraphics[width=.5\textwidth]{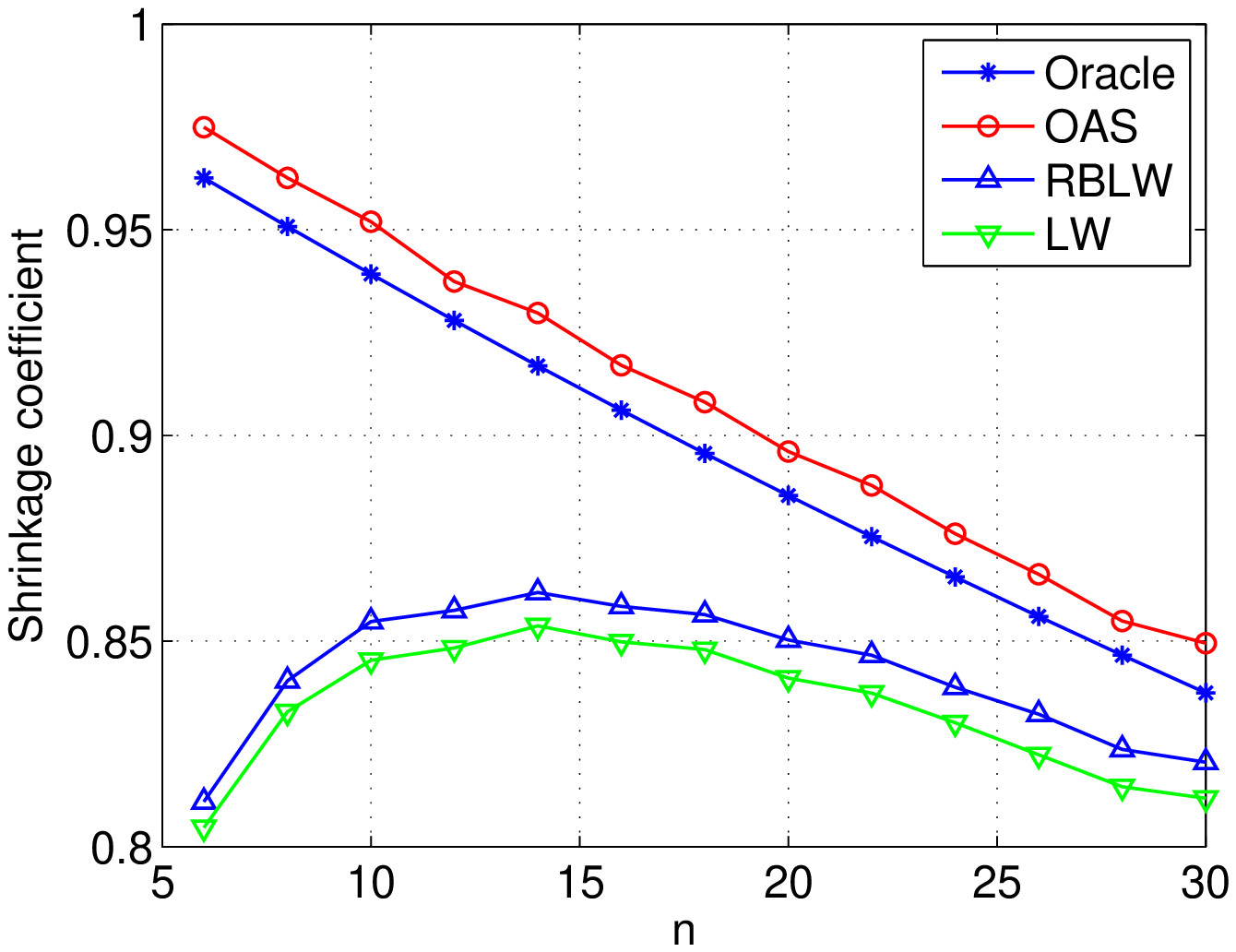}}
  \caption{AR(1) process: Comparison of covariance estimators when $p
    = 100$, $r=0.5$.}
 \label{fig:2}
\end{figure}

\begin{figure}[htbp]
  \centering
 \subfigure[MSE]{
    \label{fig:3:a} 
    \includegraphics[width=.5\textwidth]{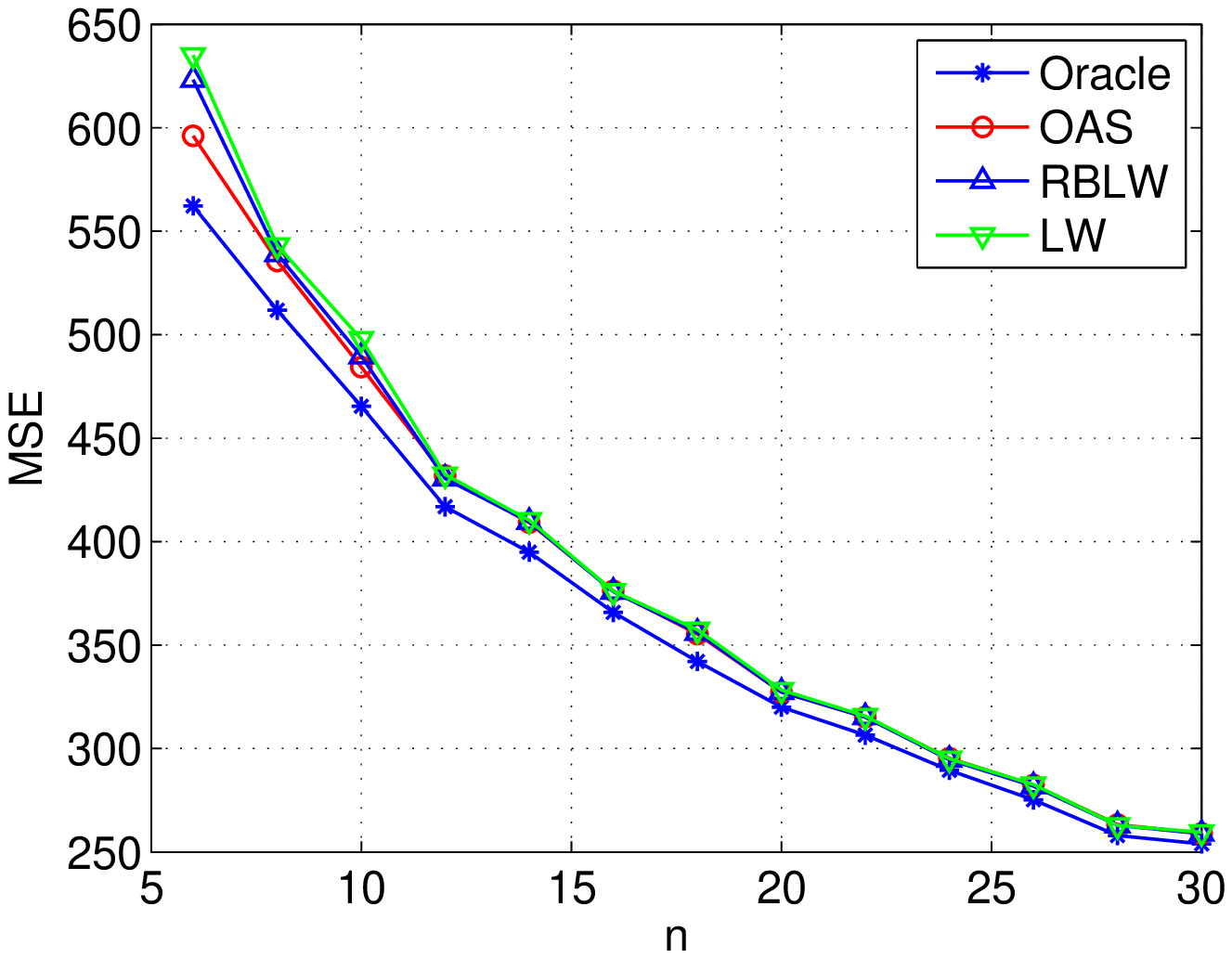}}
   \subfigure[Shrinkage coefficients]{
    \label{fig:3:b} 
    \includegraphics[width=.5\textwidth]{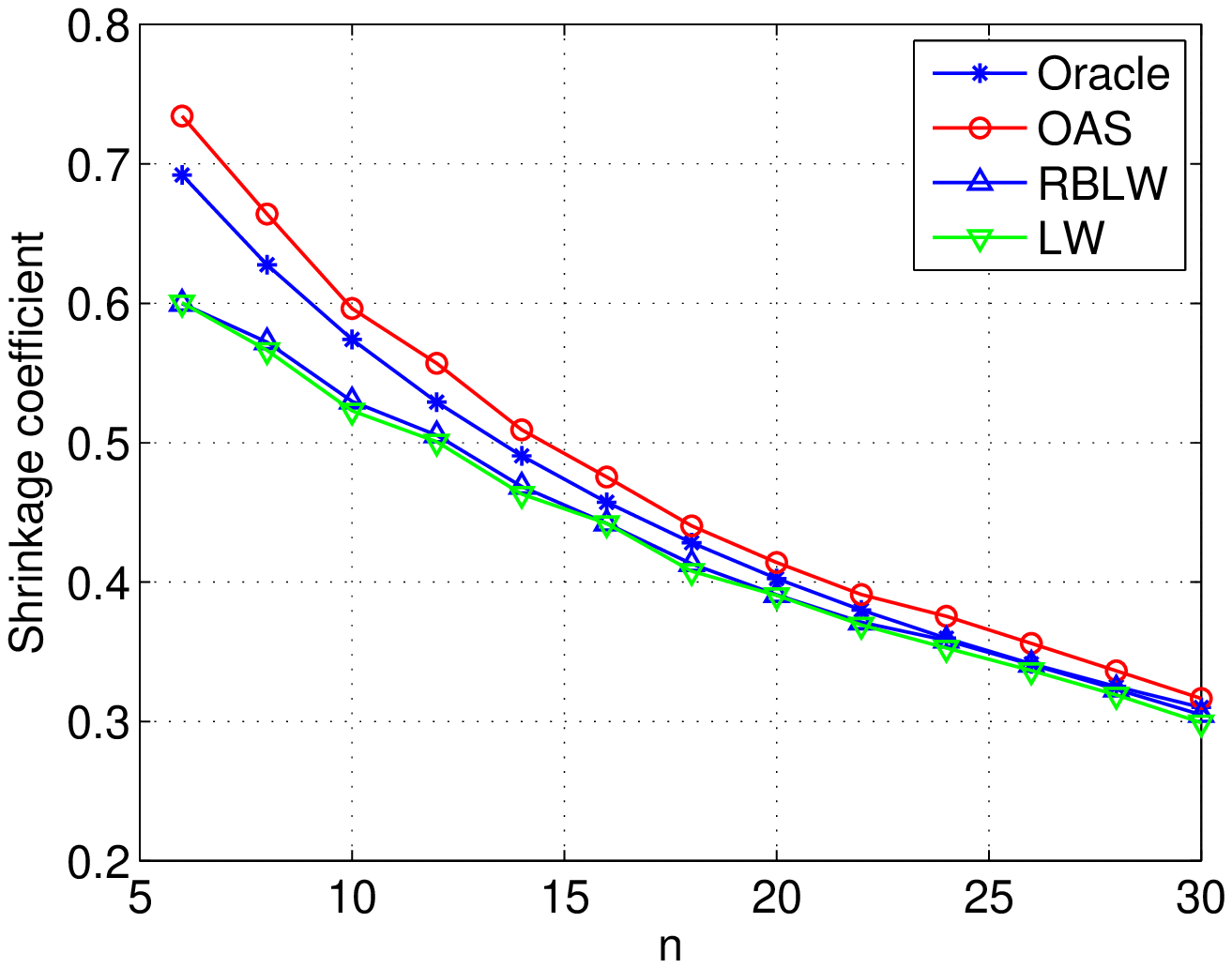}}
  \caption{AR(1) process: Comparison of covariance estimators when $p
    = 100$, $r=0.9$.}
 \label{fig:3}
\end{figure}

In the first experiment, an autoregressive covariance structured
$\bbsg$ is used. We let $\bbsg$ be the covariance matrix of a
Gaussian AR(1) process \cite{Pandit},
\begin{equation}
  \label{eq:sigma_AR_1}
\bbsg_{ij} = r^{|i-j|},
\end{equation}
where $\bbsg_{ij}$ denotes the entry of $\bbsg$ in row $i$ and
column $j$. We take $r = 0.1, 0.5 ~\text{and}~ 0.9$ for the
different simulations reported below. Figs.
\ref{fig:1:a}-\ref{fig:3:a} show the MSE of the estimators for
different values of $r$. Figs. \ref{fig:1:b}-\ref{fig:3:b} show
the corresponding shrinkage coefficients.

\begin{figure}[htbp]
  \centering
   \includegraphics[angle=0,width=.5\textwidth]{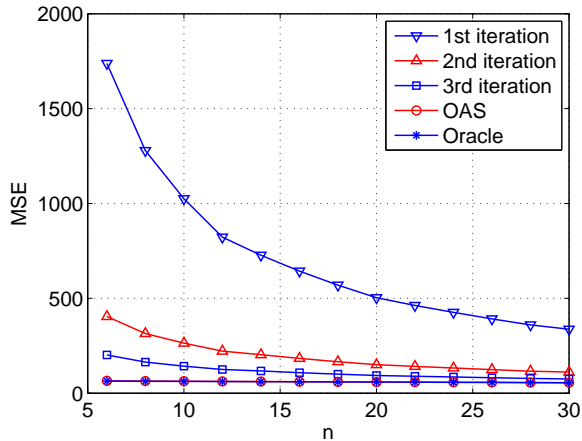}
   \caption{AR(1) process: Comparison of MSE in different iterations,
     when $p=100$, $r=0.5$}
   \label{fig:iterations}
\end{figure}

In Fig. \ref{fig:iterations} we plot the MSE of the first three
iterations obtained by the iterative procedure in
(\ref{eq:iteration_Sigma}) and (\ref{eq:iteration_rho}). For
comparison we also plot the results of the OAS and the oracle
estimator. We set $r=0.5$ in this example and start the iterations
with the initial guess $\hsg_0 = \hs$. From Fig.
\ref{fig:iterations} it can be seen that as the iterations
proceed, the MSE gradually decreases towards that of the OAS
estimator, which is very close to that of the oracle.

\begin{figure}[htbp]
  \centering
 \subfigure[MSE]{
    \label{fig:4:a} 
    \includegraphics[width=.5\textwidth]{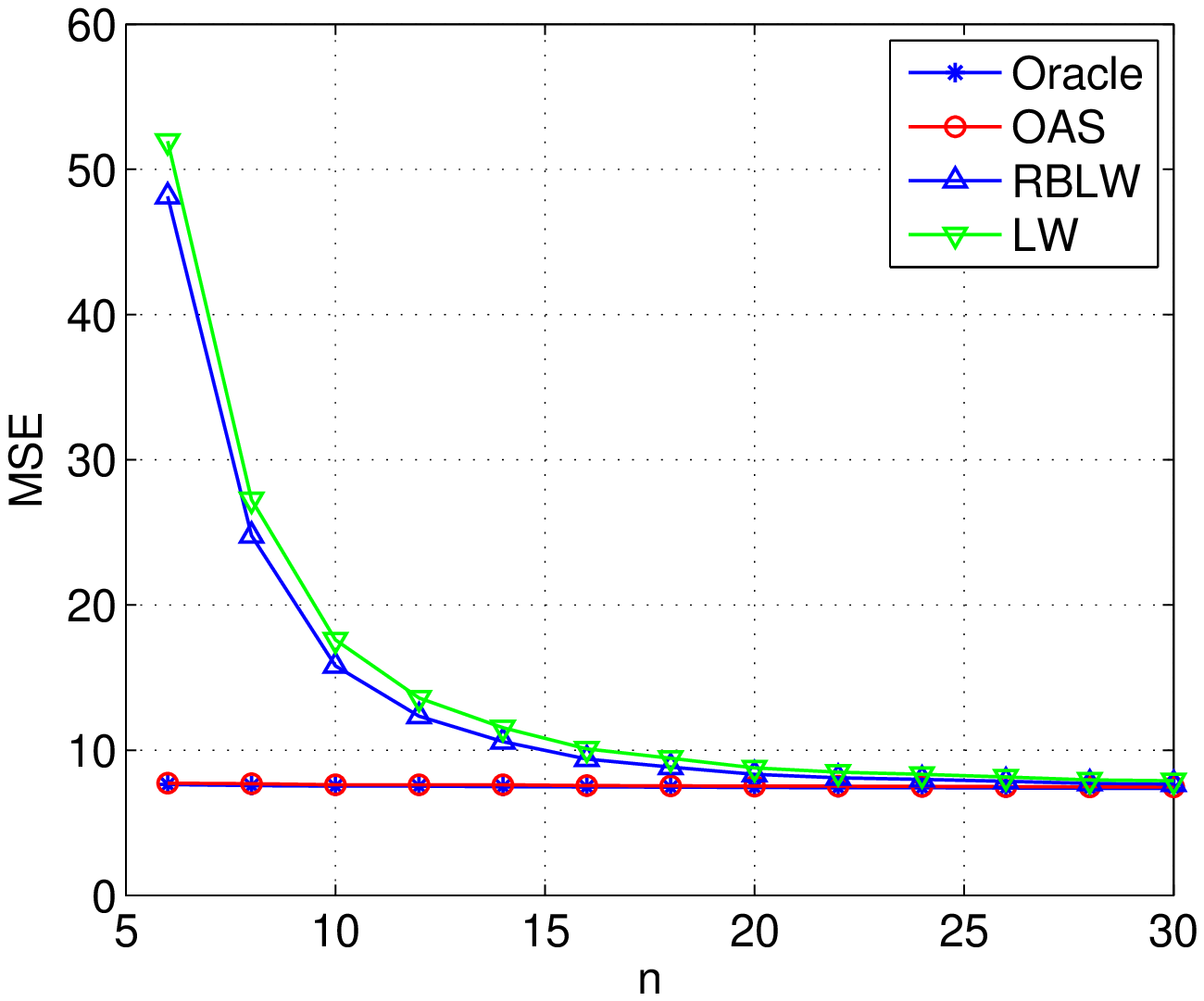}}
   \subfigure[Shrinkage coefficients]{
    \label{fig:4:b} 
    \includegraphics[width=.5\textwidth]{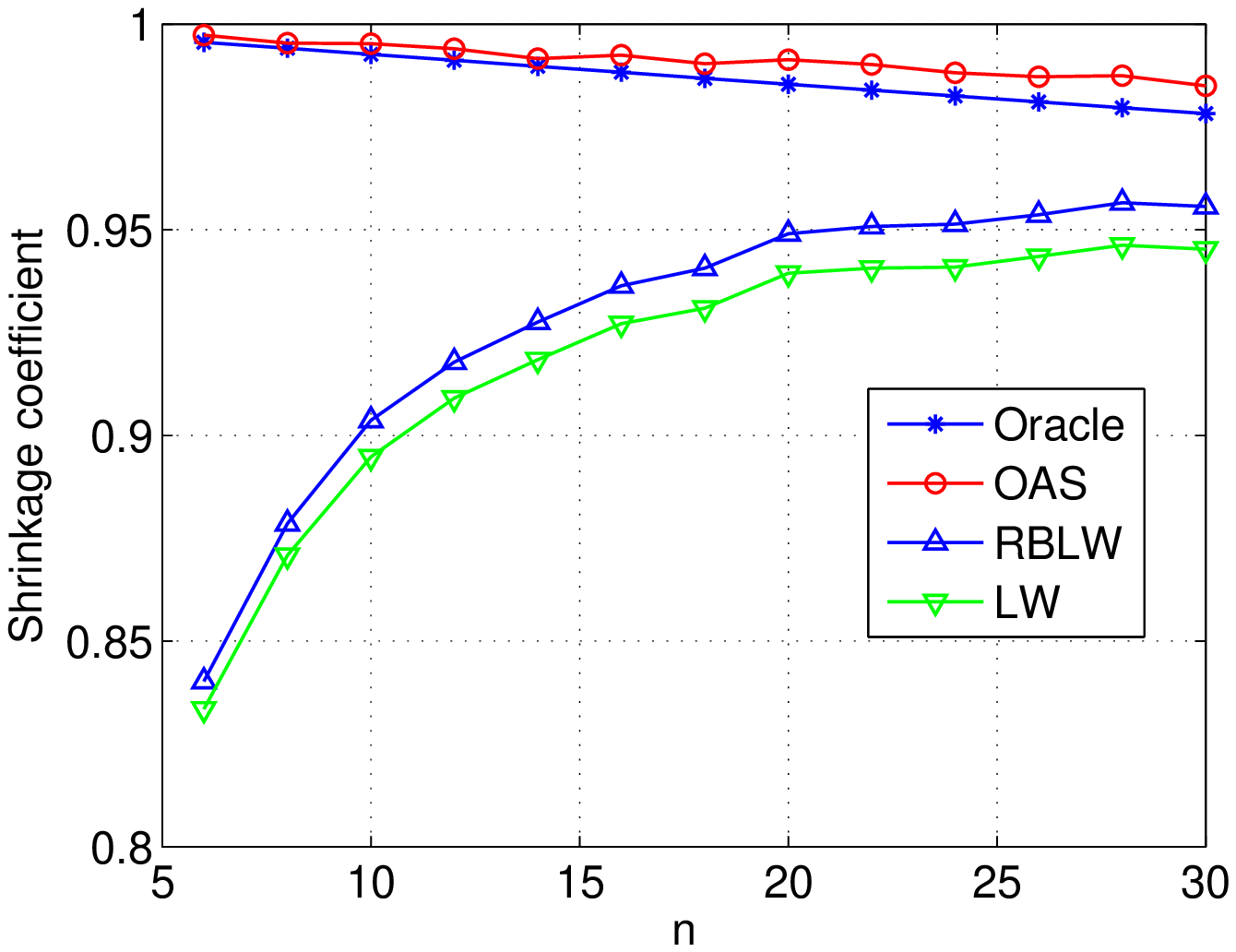}}
  \caption{Incremental FBM process: Comparison of covariance estimators when $p
    = 100$, $H=0.6$.}
 \label{fig:4}
\end{figure}

\begin{figure}[htbp]
  \centering
 \subfigure[MSE]{
    \label{fig:5:a} 
    \includegraphics[width=.5\textwidth]{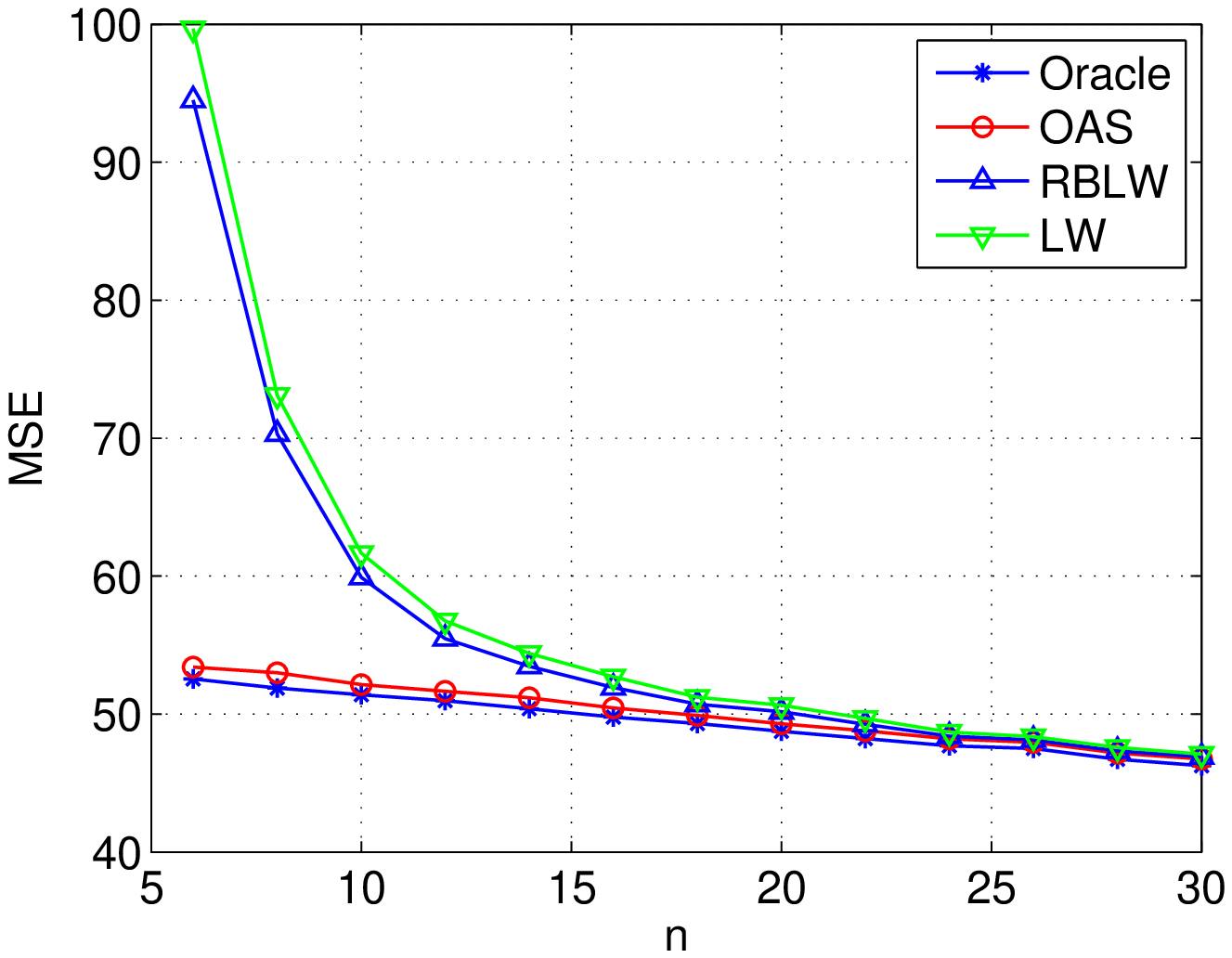}}
   \subfigure[Shrinkage coefficients]{
    \label{fig:5:b} 
    \includegraphics[width=.5\textwidth]{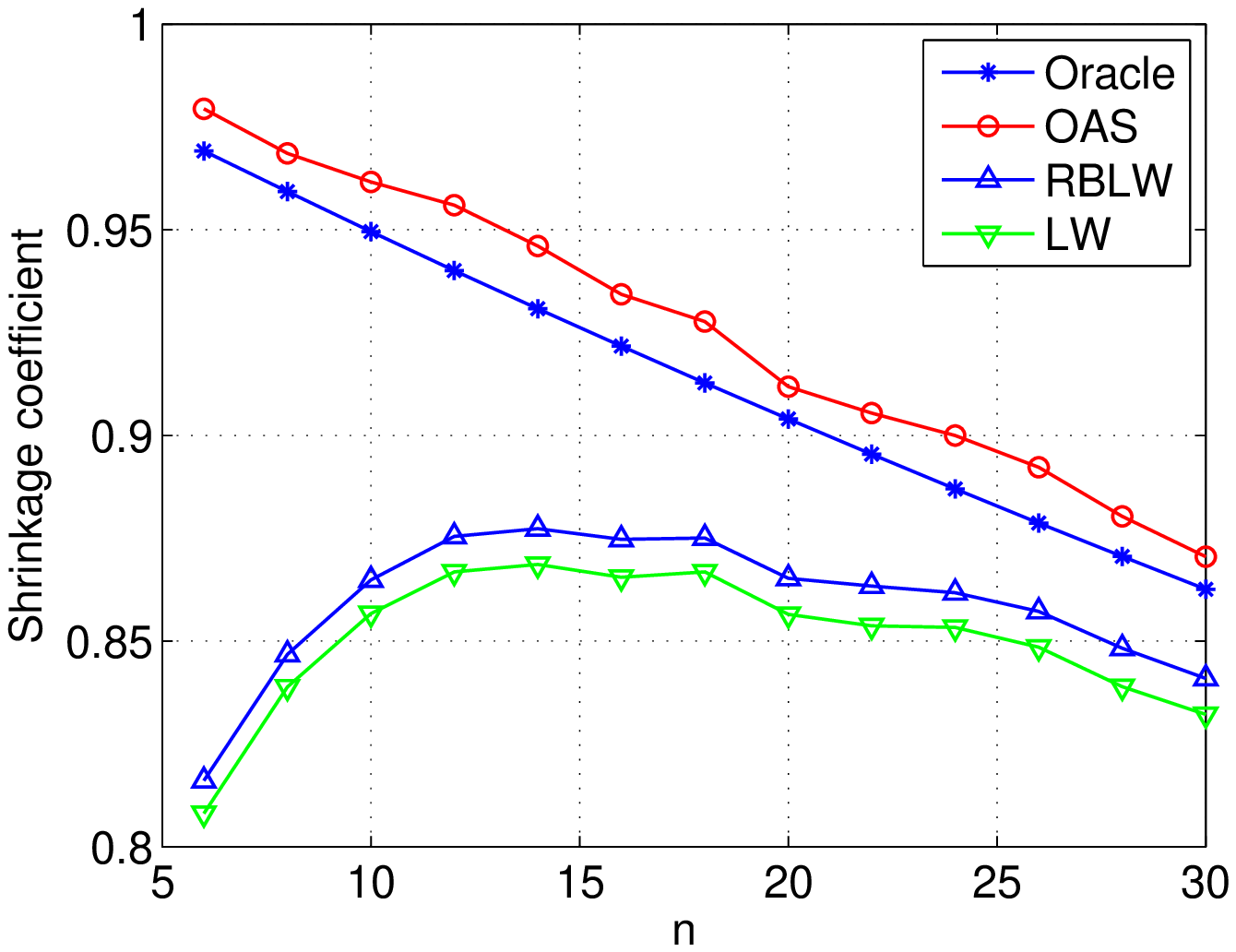}}
  \caption{Incremental FBM process: Comparison of covariance estimators when $p
    = 100$, $H=0.7$.}
 \label{fig:5}
\end{figure}

\begin{figure}[htbp]
  \centering
 \subfigure[MSE]{
    \label{fig:6:a} 
    \includegraphics[width=.5\textwidth]{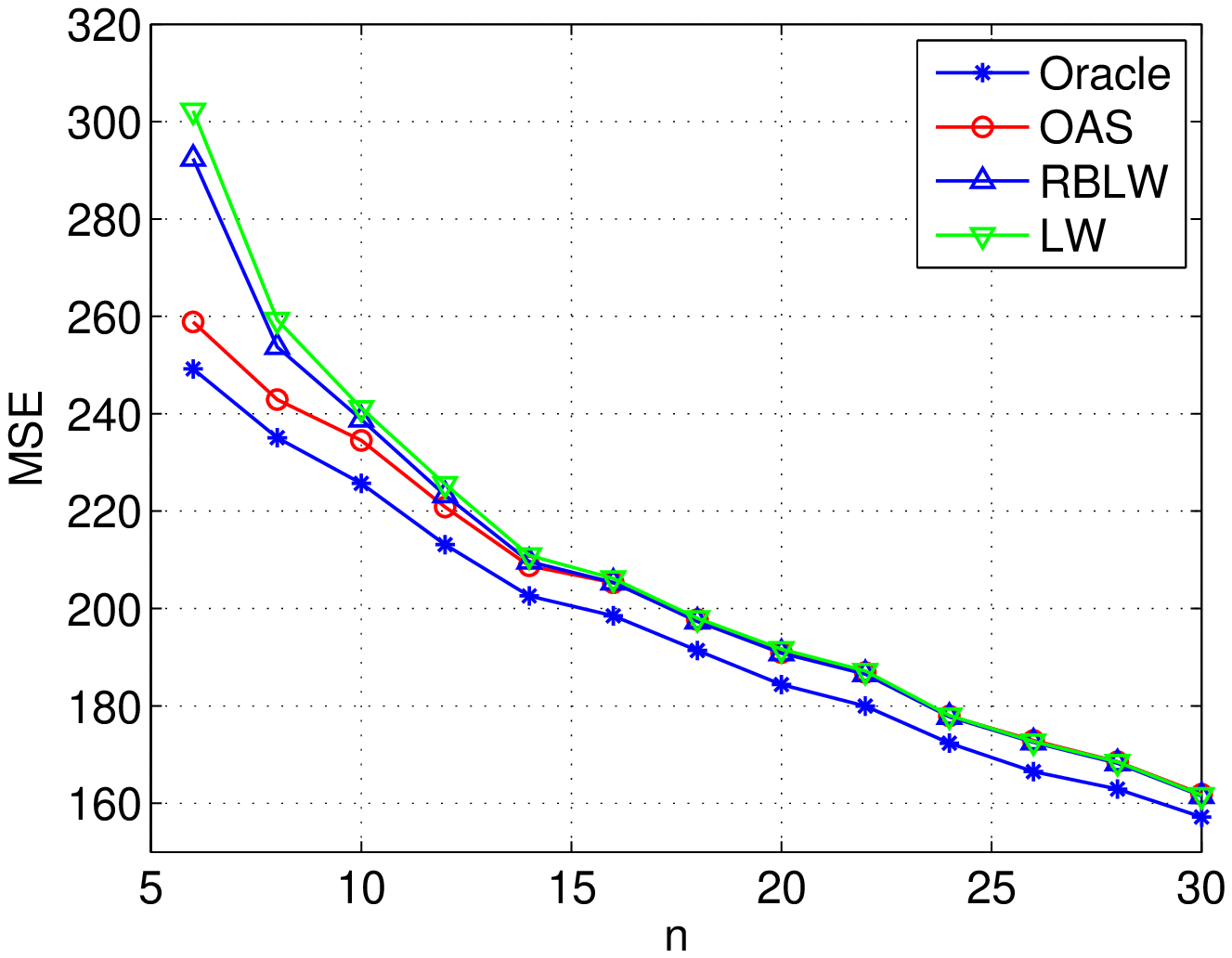}}
   \subfigure[Shrinkage coefficients]{
    \label{fig:6:b} 
    \includegraphics[width=.5\textwidth]{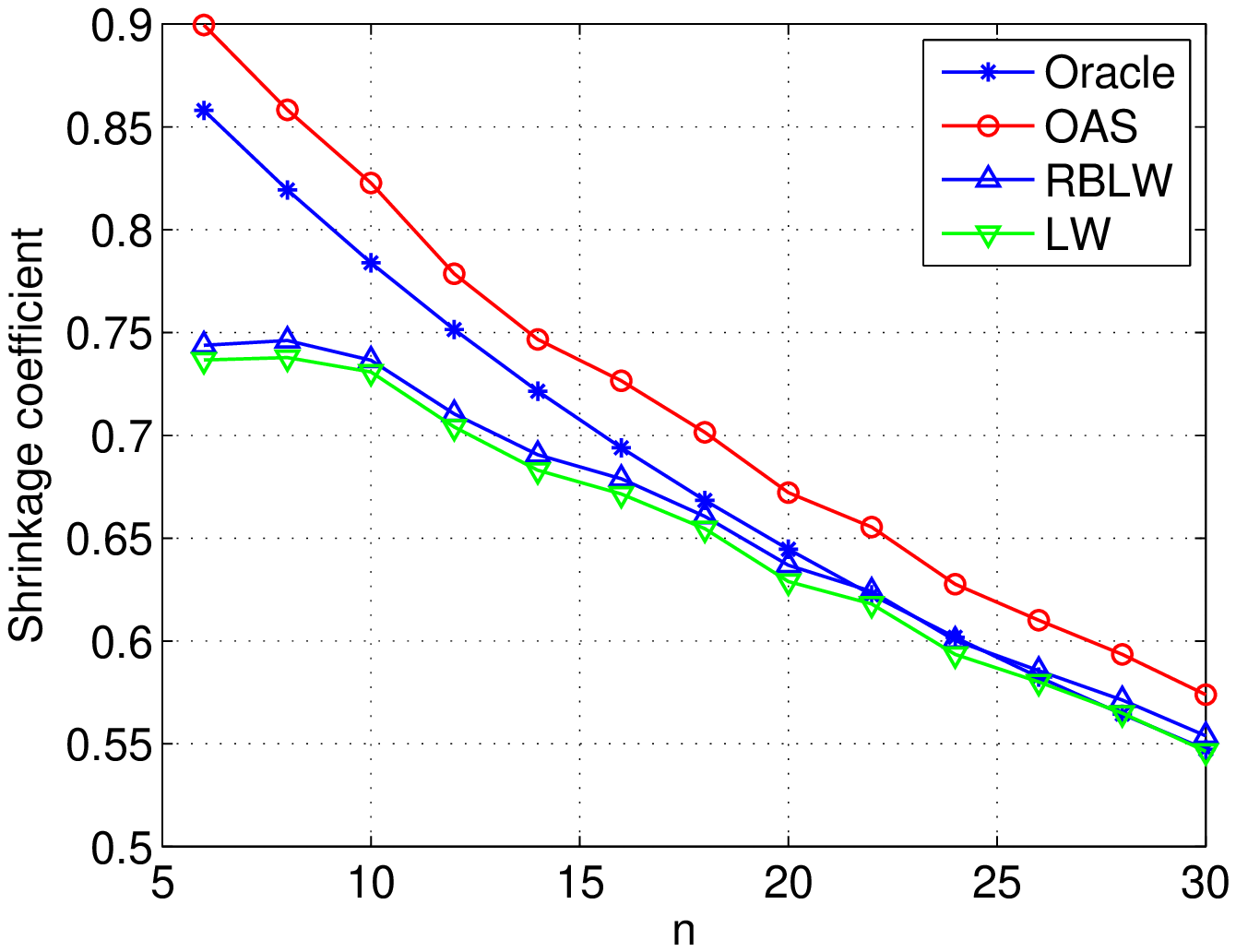}}
  \caption{Incremental FBM process: Comparison of covariance estimators when $p
    = 100$, $H=0.8$.}
 \label{fig:6}
\end{figure}

In the second experiment, we set $\bbsg$ as the covariance matrix
associated with the increment process of fractional Brownian
motion (FBM) exhibiting long-range dependence. Such processes are
often used to model internet traffic \cite{Leland} and other
complex phenomena. The form of the covariance matrix is given by
\begin{equation}
    \label{eq:FBM}
\bbsg_{ij} =
\frac{1}{2}\left[(|i-j|+1)^{2H}-2|i-j|^{2H}+(|i-j|-1)^{2H}\right],
\end{equation}
where $H \in [0.5,1]$ is the so-called Hurst parameter. The
typical value of $H$ is below 0.9 in practical applications. We
choose $H$ equal to 0.6, 0.7 and 0.8. The MSE and shrinkage
coefficients are plotted in Figs. \ref{fig:4:a}-\ref{fig:6:a} and
Figs. \ref{fig:4:b}-\ref{fig:6:b}, respectively.

From the simulation results in the above two experiments, it is
evident that the OAS estimator performs very closely to the ideal
oracle estimator. When $n$ is small, the OAS significantly
outperforms the LW and the RBLW. The RBLW improves slightly upon
the LW, but this is not obvious at the scale of the plots shown in
the figures. As expected, all the estimators converge to a common
value when $n$ increases.

As indicated in (\ref{eq:oracle_def}) and shown from simulation
results, the oracle shrinkage coefficient $\rho_O$ decreases in
the sample number $n$. This makes sense since $(1-\rho_O)$ can be
regarded as a measure of ``confidence'' assigned to $\hs$.
Intuitively, as more observations are available, one acquires
higher confidence in the sample covariance $\hs$ and therefore
$\rho_O$ decreases. This characteristic is exhibited by $\hat \rho
_{OAS}^*$ but not by $\hat\rho_{RBLW}^*$ and $\hat \rho _{LW}^*$.
This may partially explain why OAS outperforms RBLW and LW for
small samples. All the estimators perform better when the
sphericity of $\bbsg$ increases, which corresponds to small values
of $r$ and $H$.


Our experience through numerous simulations with arbitrary
parameters suggests that in practice the OAS is preferable to the
RBLW. However, as the RBLW is provably better than the LW there
exists counter examples. For the incremental FBM covariance
$\bbsg$ in (\ref{eq:FBM}), we set $H =0.9, n = 20, p = 100$. The
simulation is repeated for 5000 times and the result is shown in
Table 1, where MSE($\hsg_{RBLW}$) $<$ MSE($\hsg_{OAS}$) $<$
MSE($\hsg_{LW}$). The differences are very small but establish
that the OAS estimator does not always dominate the RBLW. However,
we suspect that this will only occur when $\bbsg$ has a very small
sphericity, a case of less interest in practice as small
sphericity of $\bbsg$ would suggest a different shrinkage target
than $\hbbF$.
\begin{table}
  \centering
  \caption{Incremental FRM process: comparison of MSE and shrinkage coefficients when $H = 0.9, n = 20, p = 100$.}\label{table:1}
  \begin{tabular}{|c|c|c|c|c|}
  \hline
   & MSE & Shrinkage coefficient \\
   \hline
   Oracle & 428.9972 & 0.2675 \\
   \hline
   OAS & 475.2691    & 0.3043 \\
   \hline
   RBLW   & 472.8206    & 0.2856 \\
   \hline
   LW  & 475.5840    & 0.2867 \\
  \hline
\end{tabular}
\end{table}

\subsection{Application to the Capon beamformer}
Next we applied the proposed shrinkage estimators to the signal
processing problem of adaptive beamforming. Assume that a
narrow-band signal of interest $s(t)$ impinges on an unperturbed
uniform linear array (ULA) \cite{Stoica2005} comprised of $p$
sensors. The complex valued vector of $n$ snapshots of the array
output is
\begin{equation}
\label{eq:beamforming} \bx(t) = \ba(\theta_s)s(t) + \bn(t), \quad
\text{for} \quad t = 1,\ldots,n,
\end{equation}
where $\theta_s$ is parameter vector determining the location of
the signal source and $\ba(\theta)$ is the array response for a
generic source location $\theta$. Specifically,
\begin{equation}
\ba(\theta) = [1, e^{-j\omega},
e^{-j2\omega},\ldots,e^{-j(p-1)\omega}]^T,
\end{equation}
where $\omega$ is the spatial frequency. The noise/interference
vector $\bn(t)$ is assumed to be zero mean i.i.d. Gaussian
distributed. We model the unknown $s(t)$ as a zero mean i.i.d.
Gaussian process.

In order to recover the unknown $s(t)$ the Capon beamformer
\cite{Stoica2005} linearly combines the array output $\bx(t)$
using a vector of weights $\vec w$, calculated by
\begin{equation}
\label{eq:capon} \vec w =
\frac{\bbsg^{-1}\ba(\theta_s)}{\ba(\theta_s)^H \bbsg^{-1}
\ba(\theta_s)},
\end{equation}
where $\bbsg$ is the covariance of $\bx(t)$. The covariance
$\bbsg$ is unknown while the array response $\ba(\theta)$ and the
source direction-of-arrival (DOA) $\theta_s$ are known. After
obtaining the weight vector $\vec w$, the signal of interest
$s(t)$ is estimated by $\vec w^H \vec x(t)$.

To implement (\ref{eq:capon}) the matrix $\bbsg$ needs to be
estimated. In \cite{Abrahamsson2007} it was shown  that using the
LW estimator could substantially improve Capon beamformer
performance over conventional methods. As we will see below, the
OAS and the RBLW shrinkage estimators can yield even better
results.

Note that the signal and the noise processes are complex valued
and $\bbsg$ is thus a complex (Hermitian symmetric) covariance
matrix. To apply the OAS and RBLW estimators we use the same
approach as used in \cite{Abrahamsson2007} to extend the real LW
covariance estimator to the complex case. Given a $p \times 1$
complex random vector $\bx$, we represent it as a $2p \times 1$
vector of its real and imaginary parts
\begin{equation}
\label{eq:stacked_vec_def} \bx_{s} = \(
\begin{aligned}
\text{Re}\(\bx\)\\
\text{Im}\(\bx\)
\end{aligned}
\).
\end{equation}
Then the estimate of the complex covariance can be represented as
\begin{equation}
\label{eq:cov_real} \hsg_s = \bl
\begin{aligned}
& \hsg_{rr} &\hsg_{ri}\\
& \hsg_{ir} &\hsg_{ii}\\
\end{aligned}
\br
\end{equation}
where $\hsg_{rr}$, $\hsg_{ri}$, $\hsg_{ir}$ and $\hsg_{ii}$ are $p
\times p$ sub-matrices. The real representation
(\ref{eq:cov_real}) can be mapped to the full complex covariance
matrix $\bbsg$ as
\begin{equation}
\hsg = \bl\hsg_{rr}+\hsg_{ii}\br + j\bl\hsg_{ir}-\hsg_{ri}\br.
\end{equation}
Using this representation we can easily extend the real valued LW,
RBLW and OAS estimators to complex scenarios.

We conduct the beamforming simulation as follows. A ULA of $p =
10$ sensor elements with half wavelength spacing is assumed and
three signals were simulated as impinging on the array. The signal
of interest has a DOA $\theta_s = 20^\circ$ and a power
$\sigma_s^2 =$ 10 dB above the complex Gaussian sensor noise. The
other two signals are mutually independent interferences. One is
at DOA angle of $\theta_{i1} = -30^\circ$ and the other one is
close to the source of interest with its angular location
corresponding to a spatial frequency of
\[
\omega_{i2} = \pi\sin(\theta_s) + 2\pi \frac{\gamma}{p}
\]
where $\gamma$ is set to 0.9. Each signal has power 15 dB above
the sensor noise.

\begin{figure}%
\begin{center}
\includegraphics[width=.5\textwidth]{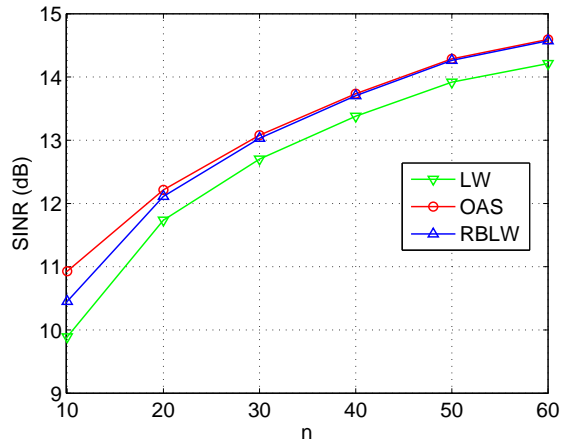}%
\caption{Comparison between different covariance shrinkage
estimators in the Capon beamformer. SINR is plotted versus number
of snapshots $n$. OAS achieves as much as 1 dB improvement over
the LW.}
\label{fig:beamforming}%
\end{center}
\end{figure}

We implemented the complex versions of the LW, the RBLW and the
OAS covariance estimators, described above, and used them in place
of $\bbsg$ in the Capon beamformer expression (\ref{eq:capon}).
The beamforming performance gain is measured by the SINR defined
as \cite{Abrahamsson2007}
\begin{equation}
\text{mean SINR} = \frac{1}{K} \sum_{k=1}^K
\frac{\sigma_s^2\l|\hat{\vec w}_k^H \vec
a\(\theta_s\)\r|^2}{\hat{\vec w}_k^H [\bbsg - \sigma_s^2 \vec
a(\theta_s) \vec a(\theta_s) ^H] \hat{\vec w}_k},
\end{equation}
where $K$ is the number of Monte-Carlo simulations and $\hat{\vec
w}_k$ is the weight vector obtained by (\ref{eq:capon}) in the
$k$th simulation. Here $K = 5000$ and $n$ varies from 10 to 60 in
step of 5 snap shots. The gain is shown in Fig.
\ref{fig:beamforming}. In \cite{Abrahamsson2007} it was reported
that the LW estimator achieves the best SINR performances among
several contemporary Capon-type beamformers. It can be seen in
Fig. \ref{fig:beamforming} that the RBLW and the OAS do even
better, improving upon the LW estimator. Note also that the
greatest improvement for OAS in the small $n$ regime is observed.

\section{Conclusion}

In this paper, we introduced two new shrinkage algorithms to
estimate covariance matrices. The RBLW estimator was shown to
improve upon the state-of-the-art LW method by virtue of the
Rao-Blackwell theorem. The OAS estimator was developed by
iterating on the optimal oracle estimate, where the limiting form
was determined analytically. The RBLW provably dominates the LW,
and the OAS empirically outperforms both the RBLW and the LW in
most experiments we have conducted. The proposed OAS and RBLW
estimators have simple explicit expressions and are easy to
implement. Furthermore, they share similar structure differing
only in the form of the shrinkage coefficients. We applied these
estimators to the Capon beamformer and obtained significant gains
in performance as compared to the LW Capon beamformer
implementation.

Through out the paper we set the shrinkage target as the scaled
identity matrix. The theory developed here can be extended to
other non-identity shrinkage targets. An interesting question for
future research is how to choose appropriate targets in specific
applications.
\section{Appendix}

In this appendix we prove Theorem \ref{thm:RBLW}. Theorem
\ref{thm:RBLW} is non-trivial and requires careful treatment using
results from the theory of Haar measure and singular Wishart
distributions. The proof will require several intermediate results
stated as lemmas. We begin with a definition.

\begin{definition}
    \label{eq:def_1}
Let $\blc\bx_i\brc_{i=1}^n$ be a sample of $p$-dimensional i.i.d.
Gaussian vectors with mean zero and covariance $\bbsg$. Define a
$p \times n$ matrix $\bbx$ as
\begin{equation}
   \label{eq:X_def}
\bbx = \bl\bx_1,\bx_2,\ldots,\bx_n\br.
\end{equation}
Denote $r = \min(p,n)$ and define the singular value decomposition
on $\bbx$ as
\begin{equation}
    \label{eq:SVDX}
\bbx = \bbh \bbla \bbq,
\end{equation}
where $\bbh$ is a $p \times r$ matrix such that $\bbh^T\bbh =
\bbi$, $\bbla$ is a $r \times r$ diagonal matrix in probability 1,
comprised of the singular values of $\bbx$, and $\bbq$ is a $r
\times n$ matrix such that $\bbq\bbq^T = \bbi$.
\end{definition}

Next we state and prove three lemmas.
\begin{lem}
  \label{lem:shanghai_1}
Let $\bl\bbh,\bbla,\bbq\br$ be matrices defined in Definition
\ref{eq:def_1}. Then $\bbq$ is independent of $\bbh$ and $\bbla$.
\end{lem}
\begin{proof}
For the case $n \le p$, $\bbh$ is a $p \times n$ matrix, $\bbla$
is a $n \times n$ square diagonal matrix and $\bbq$ is a $n \times
n$ orthogonal matrix. The pdf of $\bbx$ is
\begin{equation}
p\bl\bbx\br =
\frac{1}{(2\pi)^{pn/2}\det(\Sigma)^{n/2}}e^{-\frac{1}{2}\tr\(\bbx\bbx^T\Sigma^{-1}\)}.
\end{equation}
Since $\bbx\bbx^T = \bbh\bbla\bbla^T\bbh^T$, the joint pdf of
$\(\bbh, \bbla,\bbq\)$ is
\begin{equation}
  \label{eq:beijing_2}
\begin{aligned}
& p\(\bbh, \bbla,\bbq\) =\\
& \quad
\frac{1}{(2\pi)^{pn/2}\det(\Sigma)^{n/2}}e^{-\frac{1}{2}\tr\(\bbh\bbla\bbla^T\bbh^T\Sigma^{-1}\)}J\(\bbx
\rightarrow \bbh,\bbla,\bbq\),
\end{aligned}
\end{equation}
where $J\(\bbx\rightarrow \bbh,\bbla,\bbq\)$ is the Jacobian
converting from $\bbx$ to $\(\bbh, \bbla,\bbq\)$. According to
Lemma 2.4 of \cite{Srivastava2},
\begin{equation}
  \label{eq:beijing_1}
\begin{aligned}
& J\(\bbx\rightarrow \bbh,\bbla,\bbq\) = \\
&\quad
2^{-n}\det(\bbla)^{p-n}\prod_{j<k}^n\(\lambda_j^2-\lambda_k^2\)g_{n,p}\(\bbh\)g_{n,n}\(
\bbq\),
\end{aligned}
\end{equation}
where $\lambda_j$ denotes the $j$-th diagonal element of $\bbla$
and $g_{n,p}(\bbh)$ and $g_{n,n}(\bbq)$ are functions of $\bbh$
and $\bbq$ defined in \cite{Srivastava2}.

Substituting (\ref{eq:beijing_1}) into (\ref{eq:beijing_2}),
$p\(\bbh,\bbla,\bbq\)$ can be factorized into functions of
$\(\bbh,\bbla\)$ and $\bbq$. Therefore, $\bbq$ is independent of
$\bbh$ and $\bbla$.

Similarly, one can show that $\bbq$ is independent of $\bbh$ and
$\bbla$ when $n > p$.
\end{proof}

\begin{lem}
    \label{lem:Paris_1}
Let $\bbq$ be a matrix defined in Definition \ref{eq:def_1}.
Denote $\vec q$ as an arbitrary column vector of $\bbq$ and $q_j$
as the $j$-th element of $\vec q$. Then
\begin{equation}
  \label{eq:4th_moment_1}
E\blc q _{j}^4\brc = \frac{3}{n(n+2)}
\end{equation}
and
\begin{equation}
  \label{eq:4th_moment_2}
E\blc q _{k}^2q_{j}^2\brc = \frac{1}{n(n+2)}, \quad k \ne j.
\end{equation}
\end{lem}
\begin{proof}
The proof is different for the cases that $n \le p$ and $n > p$,
which are treated separately.

\emph{(1) Case $n \le p$:}\\
In this case, $\bbq$ is a real Haar matrix and is isotropically
distributed \cite{Marzetta, Hassibi, Eldar_}, \emph{i.e.}, for any
unitary matrices $\mathbf\Phi$ and $\mathbf\Psi$ which are
independent with $\bbq$, $\mathbf\Phi \bbq$ and
$\mathbf\bbq\mathbf\Psi$ have the same pdf of $\bbq$:
\begin{equation}
  \label{eq:beijing_3}
p(\mathbf\Phi \bbq) = p(\bbq\mathbf\Psi) = p(\bbq).
\end{equation}

Following \cite{Hiai} in the complex case, we now use
(\ref{eq:beijing_3}) to calculate the fourth order moments of
elements of $\bbq$. Since $\bbq$ and
\[
\left[ {\begin{array}{*{20}c}
   {\cos \theta } & {\sin \theta } & {} & {} & {}  \\
   { - \sin \theta } & {\cos \theta } & {} & {} & {}  \\
   {} & {} & 1 & {} & {}  \\
   {} & {} & {} &  \ddots  & {}  \\
   {} & {} & {} & {} & 1  \\
\end{array}} \right]\bbq
\]
are also identically distributed, we have
\begin{equation}
  \label{eq:beijing_4}
  \begin{aligned}
    & E\blc {\bbq_{11}^4 } \brc \\
    & = E\blc {\left( {\bbq_{11} \cos \theta  + \bbq_{21} \sin \theta }
        \right)^4 } \brc \\
    & = \cos ^4 \theta E\blc {\bbq_{11}^4 } \brc + \sin ^4 \theta
    E\blc {\bbq_{22}^4 } \brc \\
    & \quad + 6\cos ^2 \theta \sin ^2 \theta E\blc
      {\bbq_{11}^2 \bbq_{21}^2 } \brc \\
    & \quad + 2\cos ^3 \theta \sin \theta E\blc {\bbq_{11}^3 \bbq_{21} } \brc +
    2\cos \theta \sin ^3 \theta E\blc {\bbq_{11} \bbq_{21}^3 } \brc \\
 \end{aligned}
\end{equation}
By taking $\theta = -\theta$ in (\ref{eq:beijing_4}), it is easy
to see that
\[
2\cos ^3 \theta \sin \theta E\blc {\bbq_{11}^3 \bbq_{21} } \brc +
    2\cos \theta \sin ^3 \theta E\blc {\bbq_{11} \bbq_{21}^3 } \brc =
    0.
\]

The elements of $\l[\bbq_{ii}\r]$ are identically distributed. We
thus have $E\blc\bbq_{11}^4\brc = E\blc\bbq_{22}^4\brc$, and hence
\begin{equation}
\begin{aligned}
& E\blc {\bbq_{11}^4 } \brc \\
& = \left( {\cos ^4 \theta  + \sin ^4
    \theta } \right)E\blc
    {\bbq_{11}^4 } \brc + 6\cos ^2 \theta \sin ^2 \theta E\blc
    {\bbq_{11}^2 \bbq_{21}^2 } \brc.
 \end{aligned}
\end{equation}
By taking $\theta = \pi/3$,
\begin{equation}
  \label{eq:beijing_6}
  E\blc {\bbq_{11}^4 } \brc =  3E\blc {\bbq_{11}^2 \bbq_{21}^2 }
  \brc.
\end{equation}

Now we consider $E\blc\(\sum_{j=1}^n \bbq_{j1}^2\)^2\brc$. Since
$\bbq^T\bbq = \bbq\bbq^T = \bbi$, $\sum\limits_{j=1}^n \bbq_{j1}^2
= 1.$ This implies
\begin{equation}
  \label{eq:beijing_5}
\begin{aligned}
1 & = \sum\limits_{j=1}^n E\blc\bbq_{j1}^4\brc + \sum\limits_{j\ne
k}
  E\blc\bbq_{j1}^2\bbq_{k1}^2\brc\\
  & = n E\blc {\bbq_{11}^4 } \brc + n(n-1) E\blc {\bbq_{11}^2 \bbq_{21}^2 }
  \brc.
\end{aligned}
\end{equation}
Substituting (\ref{eq:beijing_6}) into (\ref{eq:beijing_5}), we
obtain that
\begin{equation}
E\blc {\bbq_{11}^4 } \brc  = \frac{3}{n(n+2)},
\end{equation}
and
\begin{equation}
E\blc {\bbq_{11}^2 }{\bbq_{21}^2} \brc = \frac{1}{n(n+2)}.
\end{equation}

It is easy to see that $E\blc q _j^4\brc = E\blc\bbq_{11}^4\brc$
and $E\blc q _j^2q_k^2\brc = E\blc\bbq_{11}^2\bbq_{21}^2\brc$.
Therefore (\ref{eq:4th_moment_1}) and (\ref{eq:4th_moment_2}) are
proved for the case of $n \le p$.

\emph{(2) Case $n > p$:}\\
The pdf of $\vec q$ can be obtained by Lemma 2.2 of
\cite{Srivastava2}
\begin{equation}
  \label{eq:q}
p(\vec q) = C_1\det\(\bbi-\vec q \vec q^T\)^{(n-p-2)/2} I(\vec q
\vec q ^T \prec \bbi),
\end{equation}
where
\begin{equation}
  C_1 = \frac{\pi^{-p/2}\Gamma\{n/2\}}{\Gamma\{(n-p)/2\}},
\end{equation}
and $I\(\cdot\)$ is the indicator function specifying the support
of $\vec q$. Eq. (\ref{eq:q}) indicates that the elements of $\vec
q$ are identically distributed. Therefore, $E\blc q _j^4\brc =
E\l[q_1^4\brc$ and $E\blc q_j^2q_k^2\brc = E\blc q_1^2q_2^2\brc$.
By the definition of expectation,
\begin{equation}
E\blc q_1^4\brc = C_1\int_{\vec q \vec q ^T
    \prec \bbi}q_1^4 \det \left( {\bbi -
      \vec q \vec q ^T}
  \right)^{(n - p - 2)/2} d\vec q,
\end{equation}
and
\begin{equation}
E\blc q_1^2q_2^2\brc = C_1\int_{\vec q \vec q ^T
    \prec \bbi}q_1^2q_2^2 \det \left( {\bbi -
      \vec q \vec q ^T}
  \right)^{(n - p - 2)/2} d\vec q.
\end{equation}
Noting that
\begin{equation}
  \vec q \vec q^T \prec \bbi \Leftrightarrow \vec q^T\vec q < 1
\end{equation}
and
\begin{equation}
\det \(\bbi-\vec q \vec q^T\) = 1-\vec q^T \vec q,
\end{equation}
we have
\begin{equation}
    \begin{split}
    E\blc q_1^4\brc & = C_1\int_{\vec q^T \vec q  < 1} {q_1^4 {(1 - \vec
    q^T
        \vec q
        )^{\frac{1}{2}(n-p-2)} } } d\vec q
    \\
    & = C_1\int_{\sum\limits_{j=1}^p q_j^2  < 1} {q_1^4 {\(1 - \sum\limits_{j=1}^p
    q_j^2\)^{\frac{1}{2}(n-p-2)} } } dq_1\ldots dq_{p}.
    \end{split}
\end{equation}
By changing variable of integration $\(q_1,q_{2},\cdots,q_{p}\)$
to $\(r,\theta_1,\theta_2,\cdots,\theta_{p-1}\)$ such that
\begin{equation}
    \label{eq:proof_lemma_Jacobian_z_r_theta}
\left\{ {\begin{array}{*{20}l}
   {q_1} & =  & {r\cos \theta _1 }  \\
   {q_{2}} & =  & {r\sin \theta _1 \cos \theta _2 }  \\
   {q_{3}} & =  & {r\sin \theta _1 \sin \theta _2 \cos \theta _3 }  \\
    \vdots & &  \vdots   \\
   {q_{p-1}} & =  & {r\sin \theta _1 \sin \theta _2  \cdots \sin \theta _{p - 2} \cos \theta _{p - 1} }  \\
   {q_{p}} & =  & {r\sin \theta _1 \sin \theta _2  \cdots \sin \theta _{p - 2} \sin \theta _{p - 1} }
\end{array}} \right.,
\end{equation}
we obtain
\begin{equation}
    \begin{aligned}
E & \blc q_1^4\brc = C_1\int_0^\pi d\theta_1 \int_0^\pi
d\theta_2\cdots \int_0^\pi d\theta_{p-2}\int_0^{2\pi}
d\theta_{p-1} \\
    & \cdot \int_0^1 r^4 \cos^4\theta_1
\(1-r^2\)^{\frac{1}{2}\(n-p-2\)} \l|\frac{\partial
\(q_1,\cdots,q_p\)}{\partial
\(r,\theta_1,\cdots,\theta_{p-1}\)}\r| dr,
    \end{aligned}
\end{equation}
where $$\l|\frac{\partial \(q_1,\cdots,q_{p}\)}{\partial
\(r,\theta_1,\cdots,\theta_{p-1}\)}\r| =
r^{p-1}\sin^{p-2}\theta_1\sin^{p-3}\theta_2\cdots\sin\theta_{p-2}$$
is the Jacobian associated with the change of variable.

Therefore,
\begin{equation}
    \label{eq:proof_lemma_int_F_1}
    \begin{aligned}
        E & \blc q_1^4\brc = C_1 \cdot \int_0^\pi \cos^4\theta_1\sin^{p-2}\theta_1
        d\theta_1 \cdot\int_0^\pi\sin^{p-3}\theta_2d\theta_2
        \\
        & \cdot \int_0^\pi\sin^{p-4}\theta_3d\theta_3
        \cdots\int_0^\pi\sin\theta_{p-2}d\theta_{p-2}\int_0^{2\pi}d\theta_{p-1} \\
        & \cdot \int_0^1
        \!\!r^{p+3}\(1-r^2\)^{\frac{1}{2}\(n-p-2\)} dr\\
        & = \frac{\pi^{-p/2}\Gamma\{n/2\}}{\Gamma\{(n-p)/2\}} \cdot
        \frac{3\pi^{\frac{1}{2}}}{4}\frac{\Gamma\{(p-1)/2\}}{\Gamma\{(p+4)/2\}}
        \cdot \pi^{\frac{1}{2}}
        \frac{\Gamma\{(p-2)/2\}}{\Gamma\{(p-1)/2\}} \\
        & \cdot \pi^{\frac{1}{2}} \frac{\Gamma\{(p-3)/2\}}{\Gamma\{(p-2)/2\}}
        \cdots \pi^{\frac{1}{2}}\frac{\Gamma\{3/2\}}{\Gamma\{5/2\}}
        \cdot \pi^{\frac{1}{2}}\frac{\Gamma\{1\}}{\Gamma\{3/2\}} \cdot 2\pi \\
        & \cdot \int_0^1
        \!\!r^{p+3}\(1-r^2\)^{\frac{1}{2}\(n-p-2\)}\!\!dr\\
        & = \frac{3}{2}\frac{\Gamma\{n/2\}}{\Gamma\{(n-p)/2\}
        \Gamma\{p/2+2\}}
        \int_0^1
        \!\!r^{p+3}\(1-r^2\)^{\frac{1}{2}\(n-p-2\)} dr\\
        & = \frac{3}{2}\frac{\Gamma\{n/2\}}{\Gamma\{(n-p)/2\}
          \Gamma\{p/2+2\}} \cdot
        \frac{1}{2}\frac{\Gamma\{(n-p)/2\}\Gamma\{p/2+2\}}{\Gamma\{n/2+2\}}\\
        & =
        \frac{3\Gamma\{n/2\}}{4\Gamma\{n/2+2\}}\\
        & =
        \frac{3}{n(n+2)}.
    \end{aligned}
\end{equation}
Similarly,
\begin{equation}
    \label{eq:lem2F2proof}
    \begin{aligned}
      E&\blc q_1^2q_2^2\brc = C_1\int_{\sum\limits_{k=1}^p q_k^2  < 1}
      {q_1^2 q_{2}^2 {\(1 - \sum\limits_{k=1}^p
          q_k^2\)^{\frac{1}{2}(n-p-2)} } } \!\!\!\!\!dq_1\ldots dq_{p} \\
      &  = C_1 \int_0^\pi d\theta_1 \int_0^\pi
      d\theta_2\cdots \int_0^\pi d\theta_{p-2}\int_0^{2\pi}
      d\theta_{p-1} \\
      & \cdot \int_0^1 \!\!r^2
      \!\cos^2\theta_1 r^2\sin^2\theta_1\!\cos^2\theta_2
      \!\! \(1-r^2\)^{\frac{1}{2}\(n-p-2\)}
      \\
      & \cdot
      \l|\frac{\partial
        \(q_1,\cdots,q_{p}\)}{\partial
        \(r,\theta_1,\cdots,\theta_{p-1}\)}\r| dr\\
      & = C_1\cdot\int_0^\pi \cos^2\theta_1 \sin^{p}\theta_1 d\theta_1\cdot
      \int_0^\pi
      \cos^2\theta_2 \sin^{p-3}\theta_2 d\theta_2\\
      & \cdot  \int_0^\pi
      \sin^{p-4}\theta_3 d\theta_3\cdot \int_0^\pi \sin^{p-5}\theta_4
      d\theta_4\cdots \int_0^\pi \sin\theta_{p-2}
      d\theta_{p-2} \\
      & \cdot \int_0^{2\pi}
      d\theta_{p-1} \cdot \int_0^1
      r^{p+3}\(1-r^2\)^{\frac{1}{2}\(n-p-2\)}dr\\
      & = \frac{\pi^{-p/2}\Gamma\{n/2\}}{\Gamma\{(n-p)/2\}}
      \cdot
      \frac{\pi^{\frac{1}{2}}}{2}\frac{\Gamma\{(p+1)/2\}}{\Gamma\{p/2+2\}}\cdot
      \frac{\pi^{\frac{1}{2}}}{2}\frac{\Gamma\{(p-2)/2\}}{\Gamma\{(p+1)/2\}}\\
      &  \cdot \pi^{\frac{1}{2}}
      \frac{\Gamma\{(p-3)/2\}}{\Gamma\{(p-2)/2\}}\cdot \pi^{\frac{1}{2}}
      \frac{\Gamma\{(p-4)/2\}}{\Gamma\{(p-3)/2\}}\cdots
      \pi^{\frac{1}{2}} \frac{\Gamma\{1\}}{\Gamma\{3/2\}} \\
      & \cdot2\pi \cdot
      \frac{1}{2}\frac{\Gamma\{(n-p)/2\}\Gamma\{p/2+2\}}{\Gamma\{n/2+2\}}\\
      & = \frac{1}{n(n+2)}.
\end{aligned}
\end{equation}

Therefore, (\ref{eq:4th_moment_1}) and (\ref{eq:4th_moment_2}) are
proved for the case when $n > p$. This completes the proof of
Lemma \ref{lem:Paris_1}.
\end{proof}

\begin{lem}
    \label{lem:1}
Let $\hs$ be the sample covariance of a set of $p$-dimensional
vectors  $\l\{\vec x_i\r\}_{i=1}^n$. If  $\l\{\vec
x_i\r\}_{i=1}^n$ are i.i.d. Gaussian vectors with covariance
$\bbsg$,
\begin{equation}
    \label{eq:lemma1}
  E\blc {\left. {\left\| {\vec x_i } \right\|_2^4 }
      \right| \hs }
  \brc = \frac{n}{{n + 2}}\left[ {2\tr {(\hs^2) } +
      \tr^2{(\hs})} \right].
\end{equation}
\end{lem}
\begin{proof}
For simplicity, we work with the scaled covariance matrix $\bbm$
defined as
\begin{equation}
\bbm = \sum_{i=1}^n \vec x_i \vec x_i^T = n \hs,
\end{equation}
and calculate $E\blc {\left. {\left\| {\vec x_i } \right\|_2^4 }
\right|\bbm} \brc$ instead of $E\blc {\left. {\left\| {\vec x_i }
\right\|_2^4 } \right|\hs} \brc$. We are then going to prove that
\begin{equation}
  \label{eq:shanghai_1}
E\blc {\left. {\left\| {\vec{x}_i } \right\|_2^4 } \right|\bbm}
\brc = \frac{1}{{n\left( {n + 2}
        \right)}}\left( {2\tr\left( {\bbm^2 } \right) + \tr^2 \left(
        \bbm
        \right)} \right).
\end{equation}
We use Lemma \ref{lem:shanghai_1} and Lemma \ref{lem:Paris_1} to
establish (\ref{eq:shanghai_1}).

Let $\bbx$ and $\bl\bbh,\bbla,\bbq\br$ be matrices defined in
Definition \ref{eq:def_1}. Let $\vec q$ be the $i$-th column of
$\bbq$ defined in Definition \ref{eq:def_1}. Then
\begin{equation}
  \vec x_i = \bbh \bbla \vec q.
\end{equation}
Let
\begin{equation}
  \bbd = \bbla^2.
\end{equation}
Then
\begin{equation}
  \bbm = \bbx\bbx^T = \bbh\bbla^2\bbh^T = \bbh \bbd \bbh^T,
\end{equation}
and
\begin{equation}
  \vec x_i^T \vec x_i = \vec q^T \bbla^T \bbh^T \bbh \bbla \vec q = \vec
  q^T \bbd \vec q.
\end{equation}
Therefore we have
\begin{equation}
  E\blc {\left. {\left\| {\vec x_i} \right\|_2^4 } \right|\bbm} \brc =
  E\blc {\left. \({\vec q^T \bbd \vec q }\)^2 \right|\bbm} \brc.
\end{equation}
According to Lemma \ref{lem:shanghai_1}, $\bbq$ is independent of
$\bbh$ and $\bbla$. Since $\vec q$ is a function of $\bbq$, $\bbm$
and $\bbd$ are functions of $\bbh$ and $\bbla$, $\vec q$ is
independent of $\bbm$ and $\bbd$.

From the law of total expectation,
\begin{equation}
  \label{eq:eq_62}
    E\blc {\left. \({\vec q^T \bbd \vec q }\)^2
        \right|\bbm} \brc = E\blc\l. E\blc {\left. \({\vec q^T \bbd
            \vec q }\)^2  \right|\bbm,\bbd} \brc \r| \bbm \brc.
\end{equation}
Expand $\vec q^T\bbd\vec q$ as
\begin{equation}
{\vec q^T\bbd\vec q}  = \sum\limits_{j = 1}^n {d_j q_j^2 },
\end{equation}
where $d_j$ is the $j$-th diagonal element of $\bbd$. Since $\vec
q$ is independent of $\bbm$ and $\bbd$, according to Lemma
\ref{lem:Paris_1},
\begin{equation}
  \label{eq:eq_63}
\begin{aligned}
 E&\blc{\left. {\left( {\vec q^T \bbd\vec q} \right)^2 } \right|\bbm,\bbd} \brc \\
  & = E\blc {\left. {\sum\limits_{j = 1}^n {d_j^2 q_j^4 }  +
          \sum\limits_{j \ne k} {d_j d_k q_j^2 q_k^2 } } \right|\bbm,\bbd}
\brc \\
 & = \sum\limits_{j = 1}^n {d_j^2 E\blc {q_j^4 } \brc + }
  \sum\limits_{j \ne k} {d_j d_k E\blc {q_j^2 q_k^2 } \brc}  \\
  & = \frac{1}{{n\left( {n + 2} \right)}}\left( {3\sum\limits_{j = 1}^n
      {d_j^2  + \sum\limits_{j \ne k} {d_j d_k } } } \right) \\
  & = \frac{1}{{n\left( {n + 2} \right)}}\left( {2\tr\left( {\bbd^2 }
      \right) + \tr^2 \left( \bbd \right)} \right).
 \end{aligned}
\end{equation}
Since $\tr\(\bbd\) = \tr\(\bbm\)$ and $\tr\(\bbd^2\) =
\tr\(\bbm^2\)$, substituting (\ref{eq:eq_63}) into
(\ref{eq:eq_62}), we have
\begin{equation}
  \begin{aligned}
     E & \blc {\left. \({\vec q^T \bbd \vec q }\)^2
        \right|\bbm} \brc\\
    & = E\blc\l.\frac{1}{{n\left( {n + 2}
        \right)}}\left( {2\tr\left( {\bbd^2 } \right) + \tr^2 \left( \bbd
        \right)} \right)  \r| \bbm \brc\\
    & = E\blc\l.\frac{1}{{n\left( {n + 2}
        \right)}}\left( {2\tr\left( {\bbm^2 } \right) + \tr^2 \left( \bbm
        \right)} \right)  \r| \bbm \brc\\
    & = \frac{1}{{n\left( {n + 2}
        \right)}}\left( {2\tr\left( {\bbm^2 } \right) + \tr^2 \left( \bbm
        \right)} \right).
  \end{aligned}
\end{equation}
\end{proof}
\noindent Lemma 3 now allows us to prove Theorem \ref{thm:RBLW}.

\subsection{Proof of Theorem \ref{thm:RBLW}}
\begin{proof}
\begin{equation}
    \begin{aligned}
        \hsg_{RBLW} = & E\blc\left. \hsg_{LW} \right| \hs\brc\\
        = & E\blc\left. \(1-\hat \rho_{LW}\) \hs +  \hat \rho_{LW} \hbbF \right| \hs \brc\\
        = &  \(1- E\blc\left. \hat \rho_{LW} \right| \hs\brc\) \hs + E\blc\left. \hat \rho_{LW} \hbbF \right| \hs
        \brc.
    \end{aligned}
\end{equation}
Therefore we obtain the shrinkage coefficient of $\hsg_{RBLW}$:
\begin{equation}
    \label{eq:proof_thm2_eq1}
    \begin{aligned}
    \hat\rho_{RBLW} = & E\blc\left. \hat \rho_{LW} \right| \hs\brc\\
    = & { \frac{{ \sum\limits_{i = 1}^n E\blc \l. {\l\|\vec x_i
\vec x_i^T
            - \hs\r\|_F^2 } \r| \hs\brc}}{{n^2 \left[\tr\(\hs^2\) -
            \tr^2\(\hs\)/p\right]
            }}}.
    \end{aligned}
\end{equation}

Note that
\begin{equation}
    \begin{aligned}
\sum\limits_{i = 1}^n & {E\blc{\left. {\l\|\vec x_i \vec x_i^T -
\hs\r\|_F^2 }
      \right|\hs} \brc}\\
      & =   \sum_{i=1}^nE\blc\left.{\l\|\vec x_i \r\|_2^4
      } \right| \hs
    \brc - n\tr (\hs^2).
    \end{aligned}
\end{equation}
From Lemma \ref{lem:1}, we have
\begin{equation}
    \label{eq:proof1}
    \begin{aligned}
\sum\limits_{i = 1}^n & {E\blc {\left. {\l\|\vec x_i \vec x_i^T -
\hs\r\|_F^2 }
      \right|\hs} \brc}\\
    & = \frac{n(n-2)}{n+2}\tr\(\hs^2\) + \frac{n^2}{n+2} \tr^2\(\hs\).
    \end{aligned}
\end{equation}
Equation (\ref{eq:rhoR_def}) is then obtained by substituting
(\ref{eq:proof1}) into (\ref{eq:proof_thm2_eq1}).
\end{proof}

\end{document}